\documentclass[sigconf]{acmart}

\acmConference[ICPC 2022]{The 30th International Conference on Program Comprehension}{May 21–22, 2022}{Pittsburgh, PA, USA}

\def \toolname {\textit{PTM4Tag }}
\def \toolnamenospace {\textit{PTM4Tag}}

\usepackage[normalem]{ulem}
\usepackage[utf8]{inputenc}
\usepackage{amsmath}  
\usepackage{graphicx}
\usepackage{comment}
\usepackage{multirow}
\usepackage{enumitem}
\usepackage{array}
\usepackage{tcolorbox}
\usepackage{xcolor}
\usepackage{balance}
\usepackage{mathtools}

\newboolean{showcomments}
\setboolean{showcomments}{true}
\ifthenelse{\boolean{showcomments}}
{ }

\AtBeginDocument{%
  \providecommand\BibTeX{{%
    \normalfont B\kern-0.5em{\scshape i\kern-0.25em b}\kern-0.8em\TeX}}}

\copyrightyear{2022} 
\acmYear{2022} 
\setcopyright{acmlicensed}\acmConference[ICPC '22]{30th International Conference on Program Comprehension}{May 16--17, 2022}{Virtual Event, USA}
\acmBooktitle{30th International Conference on Program Comprehension (ICPC '22), May 16--17, 2022, Virtual Event, USA}
\acmPrice{15.00}
\acmDOI{10.1145/3524610.3527897}
\acmISBN{978-1-4503-9298-3/22/05}
\begin{document}
\title{
\toolnamenospace: Sharpening Tag Recommendation of Stack Overflow Posts with Pre-trained Models}

\author{Junda He}
\affiliation{%
  \institution{School of Computing and Information Systems}
  \country{Singapore Management University}
}
\email{jundahe@smu.edu.sg}

\author{Bowen Xu}
\affiliation{%
  \institution{School of Computing and Information Systems}
  \country{Singapore Management University}
}
\email{bowenxu.2017@smu.edu.sg}
\authornote{Corresponding author.}

\author{Zhou Yang}
\affiliation{%
  \institution{School of Computing and Information Systems}
  \country{Singapore Management University}
}
\email{zyang@smu.edu.sg}

\author{DongGyun Han}
\affiliation{%
  \institution{School of Computing and Information Systems}
  \country{Singapore Management University}
}
\email{dhan@smu.edu.sg}

\author{Chengran Yang}
\affiliation{%
  \institution{School of Computing and Information Systems}
  \country{Singapore Management University}
}
\email{cryang@smu.edu.sg}

\author{David Lo}
\affiliation{%
  \institution{School of Computing and Information Systems}
  \country{Singapore Management University}
}
\email{davidlo@smu.edu.sg}

\begin{abstract}
Stack Overflow is often viewed as one of the most influential Software Question \& Answer (SQA) websites, containing millions of programming-related questions and answers. Tags play a critical role in efficiently structuring the contents in Stack Overflow and are vital to support a range of site operations, e.g., querying relevant contents. Poorly selected tags often introduce extra noise and redundancy, which raises problems like tag synonym and tag explosion. Thus, an \textit{automated tag recommendation technique} that can accurately recommend high-quality tags is desired to alleviate the problems mentioned above.

Inspired by the recent success of pre-trained language models (PTMs) in natural language processing (NLP), we present \toolnamenospace, a {\em tag recommendation framework} for Stack Overflow posts that utilize PTMs with a triplet architecture, which models the components of a post, i.e., Title, Description, and Code with independent language models.
To the best of our knowledge, this is the first work that leverages PTMs in the tag recommendation task of SQA sites. We comparatively evaluate the performance of \toolname based on five popular pre-trained models: BERT, RoBERTa, ALBERT, CodeBERT, and BERTOverflow. Our results show that leveraging CodeBERT, a software engineering (SE) domain-specific PTM in \toolname achieves the best performance among the five considered PTMs and outperforms the state-of-the-art Convolutional Neural Network-based approach by a large margin in terms of average $Precision@k$, $Recall@k$, and $F1$-$score@k$. We conduct an ablation study to quantify the contribution of a post's constituent components (Title, Description, and Code Snippets) to the performance of \toolnamenospace. Our results show that Title is the most important in predicting the most relevant tags, and utilizing all the components achieves the best performance.
\end{abstract}



\maketitle

\section{Introduction}
\label{sec:intro}

Software question \& answer (SQA) sites~\cite{xiaxin2013, post2vec, tagcombine, tagcnn, entagrec, entagresplusplus} have emerged as essential resources for assisting software developers. Stack Overflow (SO), one of the largest SQA sites, features a powerful platform that facilitates collaboration and communications among developers in a wide range of programming-related activities.
The site has accumulated extensive volumes of software engineering knowledge. As of January 2022, Stack Overflow has more than 17 million registered users and hosted over 22 million questions with 33 million answers.\footnote{\url{ https://stackexchange.com/sites?view=list##traffic} } 

The rapid growth of Stack Overflow highlights the need to efficiently manage the site's content at a large scale and support queries from users. A standard solution employed by modern SQA sites is to allow users to tag their posts with one or a few technical terms, which play an essential role in structuring and navigating content. 
Thus, selecting accurate and appropriate tags that summarize the question post concisely can help many aspects of the site usage, e.g., connecting the expertise among different communities, delivering the question to the appropriate set of people with the right expertise, etc. 
In the early days, Stack Overflow set no restrictions on tagging posts; users were free to create and choose any tags with arbitrary input. However, the quality of tags highly depends on users' level of expertise, English skills, writing styles and so on, making the tags selected by different users likely to be inconsistent, which triggers the tag synonym\footnote{ \url{https://stackoverflow.com/tags/synonyms}} and tag explosion~\cite{barua2014developers} problems. Such problems emphasize the desirability of a \textbf{tag recommendation technique} that learns to automatically predict tags for an unseen post based on a large volume of historical posts.

The task of tagging posts can be characterized as a \textit{multi-label classification} problem, where the goal is to select the most relevant subset of tags from a large group of tags. Tagging SO posts is considered challenging for several reasons.
As the levels of expertise vary by person, allowing users to tag their posts introduces much noise and makes the tag set rather sparse. The posts in Stack Overflow cover an extensive range of topics (e.g., over 10 thousand available tags), making it challenging to build one model to accurately capture the semantics of the post and establish connections between the posts and the related tags. 

There is a growing body of literature that focuses on the \textit{tag recommendation task} of SO posts. Recently, Xu et al.~\cite{post2vec} have proposed Post2Vec, a deep learning-based approach that has been applied to the tag recommendation task and has been shown to achieve state-of-the-art performance over a number of deep-learning-based baselines~\cite{tagcnn}. Xu et al. adopted Convolutional Neural Networks (CNNs)~\cite{schmidhuber2015deep} as the feature extractors in Post2Vec. In this work, we focus on further improving the tag recommendation performance by leveraging the transformer-based pre-trained models (PTMs), which enhanced CNN with the self-attention mechanism and pre-trained knowledge.

The transformer-based PTMs such as BERT~\cite{bert} and RoBERTa~\cite{roberta} have established great progress in the field of Natural Language Processing (NLP) and achieved phenomenal performance in several downstream tasks, i.e., Question Answering~\cite{qu2019bert}, Text Classification~\cite{jin2020bert}. Inspired by the success of PTMs in the NLP domain, there is increased interest in adapting pre-training models to the field of software engineering (SE)~\cite{CodeBERT,bertoverflow,sebert}. Such PTMs have been very effective in multi-class or pair-wise text classification tasks such as sentiment analysis~\cite{zhang2020sentiment}, API review~\cite{chengran2022saner}. However, not much SE literature has investigated the performance of PTMs in handling a \textit{multi-label classification problem} with hundreds or thousands of labels. This motivated us to explore whether BERT-based PTMs could outperform the state-of-the-art SO tag recommendation approach~\cite{post2vec}. 

Furthermore, several studies have shown that domain-specific PTMs can perform significantly better than language models that are pre-trained on the general domain text in modeling the target domain ~\cite{bertsurvey, biobert, scibert, clinicalbert}. Generally speaking, text in different domains
usually obeys a different word distribution, and general-purpose PTMs may fail to capture the correct semantics of technical jargon. To give an example in the SE domain, the word "Cookies" would refer to a small chunk of data
stored in the user end by the web browser to support easy access
to websites, and it does not refer to the "baked biscuit." However, a high-quality, large-scale corpus of a specialized domain usually is extremely difficult to obtain. General-purpose language models
like BERT~\cite{bert} are usually trained on a vast amount of data which
is usually significantly larger than domain-specific data. Considering
that each kind of model has its potential strengths and weaknesses,
it motivates us to explore the impact and limitations of adopting different PTMs.

In this work, we propose \toolnamenospace, a framework that trains a BERT-based multi-label classifier to recommend tags for SO posts. To examine the effectiveness and contain a better understanding of \toolnamenospace, we are interested in answering the following research questions:

\vspace{-2mm}
\subsubsection*{\textbf{RQ1: Out of the five variants of \toolname with different PTMs, which gives the best performance?}}

Considering that each model has its potential strengths and weaknesses, it motivates us to study the impact of adopting different PTMs in \toolnamenospace. Hence, we investigate this research question by implementing \toolname with different BERT-based models. Namely, we compare the results of domain-specific PTMs: CodeBERT~\cite{CodeBERT} and BERTOverflow~\cite{bertoverflow} with general domain PTMs: BERT~\cite{bert}, RoBERTa~\cite{roberta}, and ALBERT~\cite{albert}.

\vspace{-2mm}
\subsubsection*{\textbf{RQ2: How is the performance of \toolname compared to the state-of-the-art approach in Stack Overflow tag recommendation?}}
In this research question, we examine the effectiveness and performance of \toolname by comparing it with the CNN-based state-of-the-art baseline for the tag recommendation task of Stack Overflow, i.e., Post2Vec~\cite{post2vec}.

\vspace{-2mm}
\subsubsection*{\textbf{RQ3: Which component of post benefits \toolname the most?}}

\toolname is implemented with a triplet architecture, which encodes the three components of a SO post, i.e., Title, Description, and Code with different Transformer-based PTMS. Considering that each component may carry a different level of importance for the tag recommendation task, it inspires us to explore the contribution of each component by conducting an ablation study.

The contributions of the paper are as follows:

\begin{enumerate}[leftmargin=*]

\item To the best of our knowledge, our work is the first to leverage BERT-based pre-trained language models for tag recommendation of SQA sites.
We explore the effectiveness of different PTMs by training five variants of \toolname and compared their performance by categorizing them into two groups, generic and domain-specific.
Based on our findings, we advocate SE researchers considering CodeBERT as their first attempt or baseline approach for SQA-related tasks.

\item Our experiment results demonstrate that \toolname with CodeBERT outperforms the state-of-the-art CNN-based approach by a large margin but \toolname with ALBERT and BERTOverflow perform worse than the state-of-the-art approach. Thus, we conclude that leveraging PTM can help to achieve more promising results than the existing approach, but PTM within \toolname needs to be rationally selected.

\item The results of our ablation study show that Title is the most significant component for tag recommendation of SO. Still, considering all components of a post would yield the best performance of \toolnamenospace.
\end{enumerate}

The paper is structured as follows: Section \ref{sec:background} introduces the background knowledge of the tag recommendation task of
SO, the state-of-the-art baseline approach and five popular PTMs that are investigated in our study. Section \ref{sec:methodology} describes our proposed approach in detail, and then Section \ref{sec:experiment} specifies the experimental settings. Sections \ref{sec:results} presents the experimental results with analysis. In Section  \ref{sec:discussion}, we conducted a qualitative analysis, discussed the threats to validity as well as summarized the lessons we learned from our experiment results. Section \ref{sec:related_work} reviews the literature on PTMs applied in SE, and the SQA-oriented tag recommendation approaches. Finally, in Section \ref{sec:conclusion} we conclude our work and mention the future work.
\section{Background}
\label{sec:background}
In this section, we first formalize the task of recommending tags for SO post as a \textit{multi-label classification problem}. Then, we describe \textit{Post2Vec}~\cite{post2vec}, the current state-of-the-art \textit{tag recommendation approach}. In the end, we briefly introduce the five pre-trained language models that are investigated in this paper. 

\subsection{Tag Recommendation Problem}
\label{sec:formal}
Considering a SO post can be labeled by one or multiple tags, we regard the task of recommending tags for SO post as a \textit{multi-label classification problem}.
We assume to have a corpus of SO posts (denoted by $\mathcal{X}$) and a collection of tags (denoted by $\mathcal{Y}$). Formally speaking, given a SO post $x \in \mathcal{X}$, the tag recommendation task aims to acquire a function $f$ that maps $x$ to a subset of tags $y = \{y_1, y_2, ..., y_l\}  \subset \mathcal{Y}$ that are most relevant to the post $x$. We denote the total number of training examples as $N$, the total number of available tags\footnote{\url{https://stackoverflow.com/help/tagging}} as $L$ and the number of tags of a training data as $l$, such that $L = |\mathcal{Y}|$ and $l = | y |$. Noted that a SO post can be labeled with at most five tags, so $l$ must be no larger than 5.

\subsection{The State-of-the-art Approach}\label{sec:post2vec}
Xu et al. proposed Post2Vec~\cite{post2vec}, a deep learning-based tag recommendation approach for SO posts, and achieved the current state-of-the-art performance. Xu et al. trained several variants of Post2Vec to examine the architecture design from multiple aspects. In this paper, we select their best-performing model as our baseline model. More specifically, it leveraged CNN as the feature extractors of the post and divided the content of a post into three components, i.e., Title, Description, and Code. Each component of a post has its own component-specific vocabulary and is modeled separately with a different neural network. A major drawback of Post2Vec is that their underlying neural assigns equal contribution to the input tokens. We have addressed this limitation by leveraging the Transformer-based PTMs, which enhanced the architecture with a self-attention mechanism and pre-trained knowledge obtained from other datasets. Additionally, the importance of each component remains unknown in the experiments of Post2Vec. We have further conducted an ablation study to investigate the contribution of each component to the task.

\subsection{Pre-trained Language Models}
Recent trends in the NLP domain have led to the rapid development of transfer learning. Especially, substantial work has shown that pre-trained language models learn practical and generic language representations which could achieve outstanding performance in various downstream tasks simply by fine-tuning, i.e., without training a new model from scratch~\cite{tracebert,jin2020bert,qu2019bert}. With proper training manner, the model can effectively capture the semantics of individual words based on their surrounding context and reflect the meaning of the whole sentence. 

In this section, we describe the five popular PTMs investigated in our experiments. We first introduce three large-scale, general-purpose PTMs. Namely, they are BERT~\cite{bert}, RoBERTa~\cite{roberta} and ALBERT~\cite{albert}. Besides, we also consider two additional pre-trained language models designed for software engineering tasks (i.e., CodeBERT~\cite{CodeBERT} and BERTOverflow~\cite{bertoverflow}).

\subsubsection*{\textbf{BERT}} BERT~\cite{bert} is based on the Transformer architecture~\cite{transformer} and contains the bidirectional attention mechanism. BERT is pre-trained on a large corpus of general text data, including the entire English Wikipedia dataset and the BooksCorpus~\cite{Zhu_2015_ICCV}. It has two pre-training tasks: Masked Language Modeling (MLM) and Next Sentence Prediction (NSP). Given an input sentence where some tokens are masked out, the MLM task predicts the original tokens for those masked tokens. Given a pair of sentences, the NSP task aims to predict whether the second sentence in the pair is the subsequent sentence to the first sentence.

\subsubsection*{\textbf{RoBERTa}} RoBERTa~\cite{roberta} is mainly based on the original architecture of BERT but modifies a few key hyper-parameters. It removes the NSP task and feeds multiple consecutive sentences into the model. RoBERTa is trained with larger batch size and learning rate on a dataset that is an order of magnitude larger than the training data of BERT~\cite{bert, Zhu_2015_ICCV}. 
    
\subsubsection*{\textbf{ALBERT}} ALBERT~\cite{albert} is claimed as \textbf{A} \textbf{L}ite \textbf{BERT}. ALBERT involves two parameter reduction techniques: factorized embedding parameterization and cross-layer parameter sharing. Additionally, it replaced the NSP task used by BERT with the Sentence Order Prediction (SOP) task. By doing so, ALBERT can significantly reduce the number of model parameters and facilitate the training process without sacrificing the model performance.
\subsubsection*{\textbf{CodeBERT}} CodeBERT follows the same architectural design as RoBERTa. However, CodeBERT \cite{CodeBERT} is pre-trained on both natural language (NL) and programming language (PL) data from the CodeSearchNet database provided by~\cite{codesearchnet}. CodeBERT considers two objectives at the pre-training stage: masked language modeling (MLM) and Replaced Token Detection (RTD). The goal of the RTD task is to identify which tokens are replaced from the given input. CodeBERT uses bimodal data (NL-PL pairs) as input at the pre-training stage. Thus, it understands both forms of data. The CodeBERT model has been proven practical in various SE-related downstream tasks, such as Natural Language Code Search~\cite{zhou2021assessing, CoSQA}, program repair~\cite{mashhadi2021apply}, etc~\cite{CodeBERT}.
\subsubsection*{\textbf{BERTOverflow}} BERTOverflow is trained by Tabassum et al.~\cite{bertoverflow} as a domain-specific PTM on 152 million sentences from Stack Overflow. Notice that BERTOverflow The authors have introduced a software-related named entity recognizer (SoftNER) that combines an attention mechanism with code snippets. The model follows the same design as the BERT model with 110 million parameters.

Recently, Mosel et al.~\cite{sebert} have proposed seBERT, which is pre-trained on textual data from Stack Overflow\footnote{\url{https://cloud.google.com/bigquery/}}, GitHub\footnote{\url{https://www.gharchive.org/}}, and Jira Issues~\cite{jira}. In contrast to the PTMs as mentioned above, the seBERT model has a larger size and contains more parameters since it is implemented with the BERT$_{LARGE}$ architecture. Fine-tuning on seBERT requires a much longer training time and more GPU memory consumption. Therefore, we did not consider seBERT in our work due to restraints of limited computational resources.

\section{methodology}
\label{sec:methodology}
This section introduces our proposed tag recommendation framework for SO posts, \toolname in detail. The overall architecture of \toolname with three stages is illustrated in Figure~\ref{fig:architecture}, which are \textbf{Pre-processing, Feature Extraction, and Classification} respectively.
We transform the raw SO data into a desirable format at the pre-processing stage. Each SO post would be decomposed into three components: Title, Description, and Code snippets. Thus, \toolname is implemented with a \emph{triplet} architecture and leverages three PTMs as encoders to generate representations for each component. At the feature extraction stage, we feed the processed data into the used PTMs and represent each component as a feature vector. The obtained feature vectors are then fed to a fusion layer to construct the representation of the SO post. At the classification stage, the classifier maps the post representation to a tag vector that indicates the probability of each tag.

\subsection{Pre-processing}
\begin{figure}
	\includegraphics[width=1\linewidth]{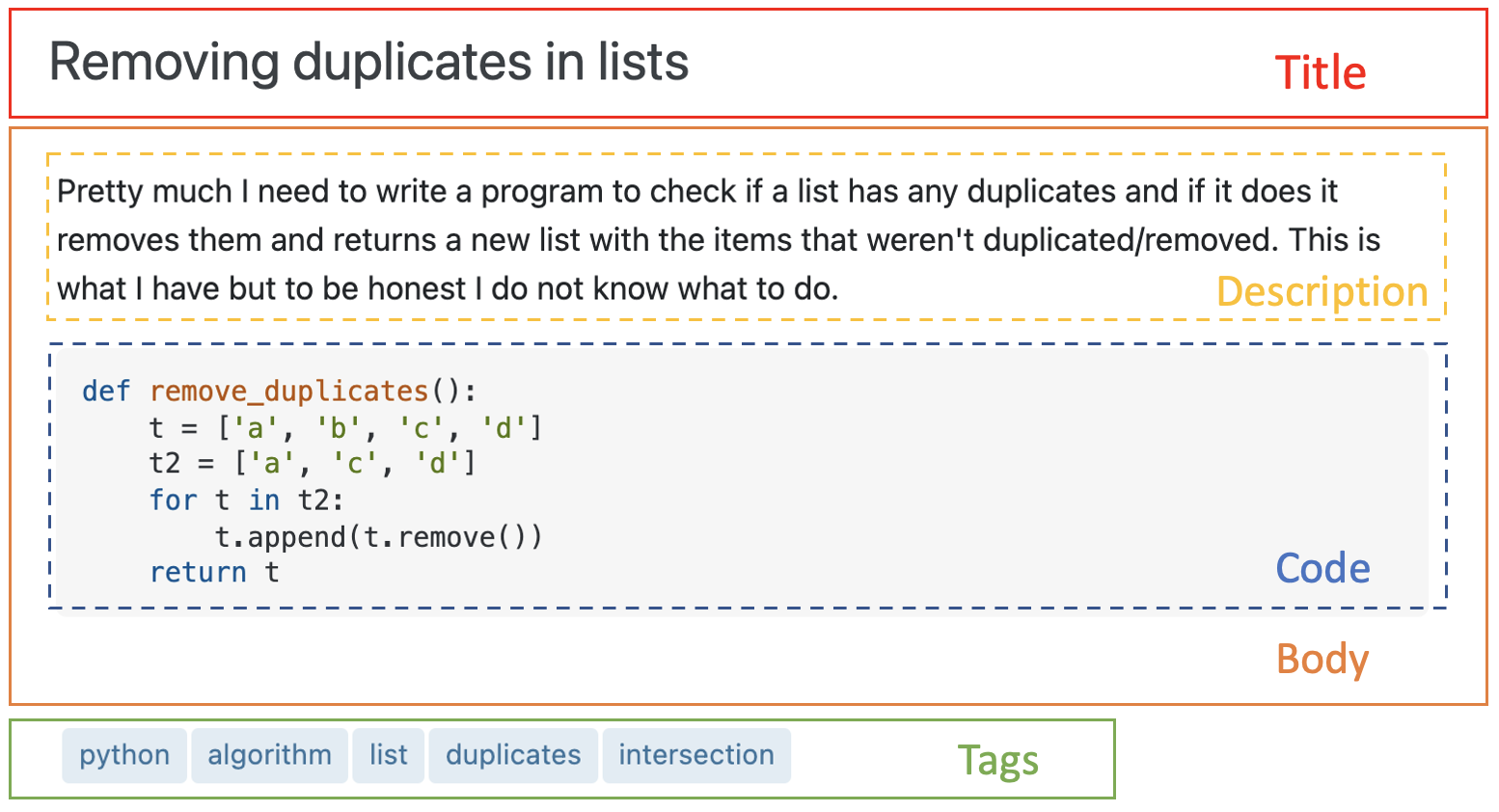}
	\vspace{-2mm}
	\caption{An example of an SO Post. A post contains a short title that summarizes the main content of this post. The body of a post can include detailed descriptions written in natural languages and code snippets.}
	\label{fig:post}
	\vspace{-4mm}
\end{figure}

\subsubsection{Post Component Extraction}
Figure~\ref{fig:post} illustrates a typical SO post that consists of three components: \textit{Title}, \textit{Body} and \textit{Tags}.
The \textit{Title} summarizes the question, and the \textit{Body} provides more details and contexts to help other users understand the question. The \textit{Body} of a SO post usually contains two parts: textual description and code snippets, which we refer to them as \textit{Description} and \textit{Code} in the following part of the paper. 

Some researchers~\cite{tagdc, tagcnn, tagcombine} consider the \textit{Code} in an SO post to be of low quality. The intuition is that users typically write the post to seek help, which means that the code snippets within the post are likely to be written by novices and have low quality. Besides, code snippets in different posts can cover a wide range of programming languages and frameworks, making it challenging to process them properly. As a result, many previous works~\cite{tagdc, tagcnn, tagcombine} have treated the \textit{Code} as noise and discarded it during the pre-processing stage. 
However, we observe that the \textit{Code} in an SO post provides valuable semantic information that can be leveraged to help in recommending tags more accurately. Take a SO post shown in Figure~\ref{fig:post} as an example, one of its tags is `{\tt python}', but neither \textit{Title} nor \textit{Description} explicitly mentions the post is {\tt python}-related. If we only look at the title and description sections, it is unclear which programming language this post is asking about. However, by only considering the \textit{Code}, the grammar of Python can be easily used to infer that the post is likely to relate to the tag `{\tt python}'.
Motivated by this example, we further divide the \textit{Body} into \textit{Description} and \textit{Code},  where the \textit{Description} corresponds to the part of \textit{Body} mainly written in natural languages and \textit{Code} refers to the code snippets within the \textit{Body}. 

According to the web design of Stack Overflow, the code snippets contained within the \textit{Body} of a SO post are usually wrapped with HTML tags \textit{(<pre><code> <\textbackslash code><\textbackslash pre>)}. Thus, we first use a regular expression formula "<pre><code>([\textbackslash s\textbackslash S]*?)<\textbackslash \textbackslash code><\textbackslash \textbackslash pre>" to split the \textit{Body} of a SO post into sections of natural language and programming language. We then further remove the redundant HTML tags within these sections since these HTML tags are used for formatting and are irrelevant to the content of a post.
By the above mean, we extract the \textit{Description} and \textit{Code} from each SO post.
Following the above procedure, we consider that a post is made up of four constituents: the \textit{Title}, \textit{Description}, \textit{Code}, and \textit{Tags}. Our proposed \toolname framework takes \textit{Title}, \textit{Description} and \textit{Code} as input and aims to predict the sets of tags that are most relevant to this post (i.e., the \textit{Tags}).

\subsubsection{PTM-Oriented Tokenization}
Since the design of \toolname leverages BERT-based PTMs, we rely on the corresponding tokenizer of the underlying pre-trained model to
generate token sequences. The BERT-based PTMs usually accept a maximum input sequence length of 512 sub-tokens which also always include two special tokens $<CLS>$ and $SEP$. The $<CLS>$ token (short for \textbf{CL}a\textbf{S}sification) is the first token of every input sequence, and the corresponding hidden state is then used as the aggregate sequence representation. $<SEP>$ token (short for \textbf{SEP}arator) is  inserted at the end of every input sequence. The problem of capturing long text arises since a significant proportion of the training point has exceeded the maximum acceptably input limit of the PTMs. According to our dataset statistics, more than 50\% of the posts' descriptions and more than 40\% of the posts' code snippets are longer than the given length limit. We tackle the problem using a head-only truncation strategy~\cite{howtofineture}, which only considers the initial 510 tokens (excluding the $<CLS>$ and $<SEP>$ tokens) as the input tokens. There are other truncation strategies such as tail-only truncation, head+tail truncation \cite{howtofineture} and other standard techniques to handle longer text such as compression \cite{howtofineture}, hierarchical method \cite{howtofineture}. However, such approaches are not attempted in our work since there is no clear conclusion on which approach would give the best performance in general, and the performance varies among different situations. We leave the extensive exploration of the effect of such methods for future work.

\subsection{PTM-based Feature Extraction}

\begin{figure}
	\centering
	\includegraphics[width=0.95\columnwidth]{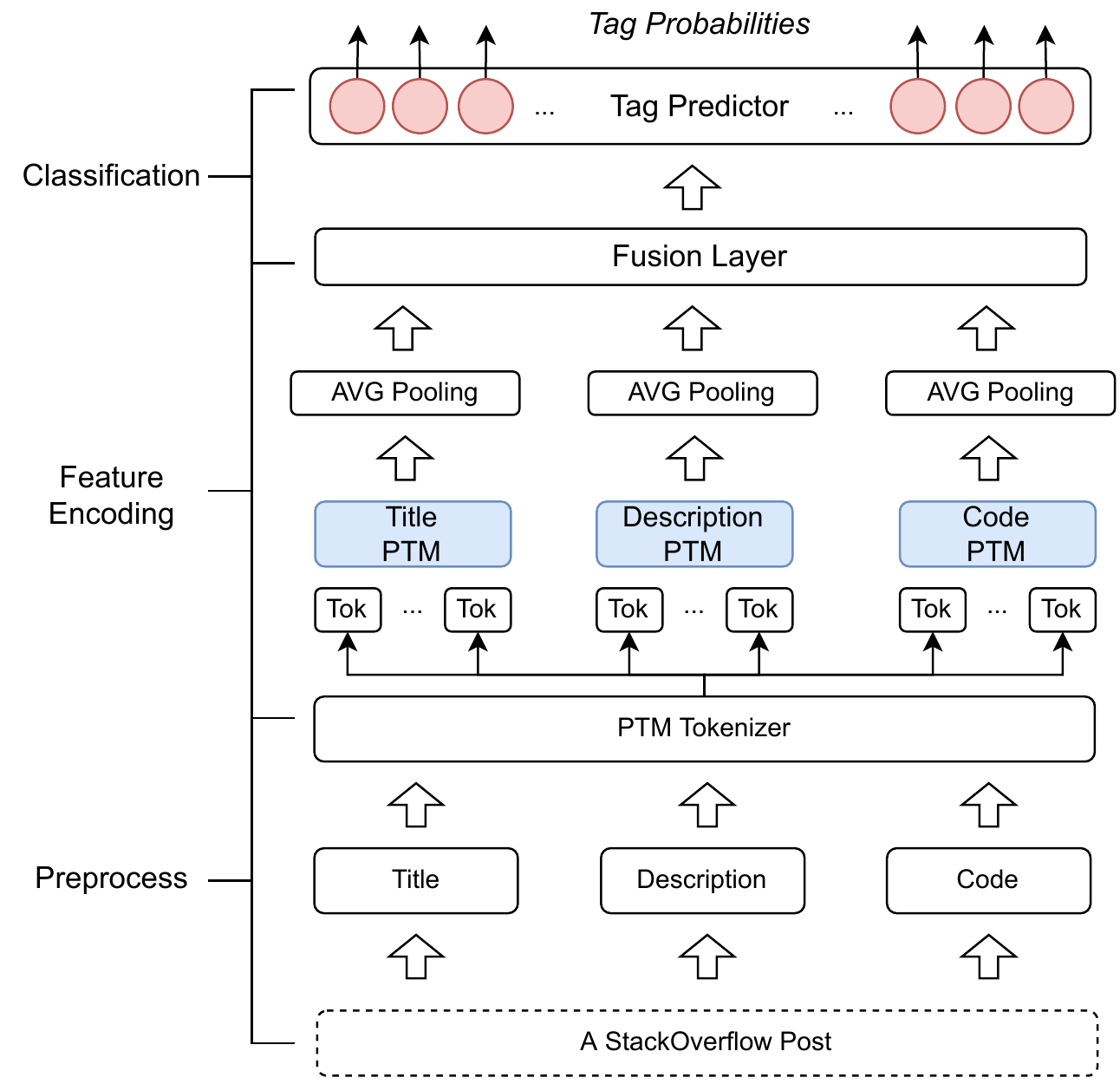}
	\caption{The overview of the \toolname framework. The title, description, and code are extracted from an SO post and fed into three different pre-trained models to obtain embeddings for each of them. A classification model takes the processed embeddings as input and produces probabilities for each tag.}
	\label{fig:architecture}
\end{figure}

\subsubsection{Language Modeling with PTMs}
A language model learns the probability distribution of sequences of words. Recently, the BERT-based PTMs have been achieving outstanding results in a range of natural language understanding tasks~\cite{jin2020bert,qu2019bert}. 
The BERT architecture~\cite{bert, transformer}, which is also inherited in other models, e.g., RoBERTa and CodeBERT is essentially the Encoder stack of transformer~\cite{transformer}. It is implemented with the self-attention mechanism~\cite{transformer} which significantly enhances its ability in capturing long dependencies. The BERT$_{BASE}$ model contains 12 layers and 110M parameters in total with 512 hidden units and 8 attention heads. BERT-based models usually are pre-trained on a large-scale corpus in order to obtain a generalized representation. Taking the expensive computational cost at the pre-training stage into account, we leveraged the released PTMs by the community in the design of \toolname to generate contextual word embeddings. The encoder part of \toolname is replaceable, and we have empirically implemented five variants of \toolname with different PTMs (refer to Section~\ref{sec:implementation}) and investigate the impact of the PTM selection within \toolname (refer to our first research question, i.e., RQ1 in Section~\ref{sec:results}).

\subsubsection{Pooling and Fusion}
After the word embeddings are generated, we obtain the post representation by applying a pooling strategy. A pooling strategy executes the down-sampling on the input representation and it is widely used to generate sentence embeddings from a sequence of word embeddings (wherein a BERT-based model, each word embedding has 768 dimensions). We generate post embeddings in the same way as generating sentence embeddings with BERT models.
There are several common choices to derive fixed size sentence embeddings from a BERT-based model, include (1) using the first \textit{<CLS>} token, (2) \textit{Average Pooling} and (3) \textit{Maximum Pooling}~\cite{reimers2019sentencebert}.

Reimers and Gurevych~\cite{reimers2019sentencebert} have evaluated the effectiveness of different pooling strategies on the SNLI dataset~\cite{bowman-etal-2015-large} and the Multi-Genre NLI dataset~\cite{williams-etal-2018-broad} in the sentence classification task, and the reported \textit{Average Pooling} gives the best performance in both datasets. 
Inspired by the findings, our proposed method leverages the \textit{Average Pooling} strategy on the hidden output to generate component-wise feature vectors by default. Finally, we concatenate these three vectors sequentially to obtain the final representation of an SO post.

\subsection{Model Training and Inference}
After performing average pooling, the output of the fusion layer is fed into a feed-forward neural network to perform the task of multi-label classification.
Given a training dataset consisting of $\mathcal{X}$ and corresponding ground truth tags $y$ for each $x \in \mathcal{X}$, we train a tag recommendation model $f$ by minimizing the following objective:
\vspace{-4mm}

\begin{equation}
 \mathcal{L} =  - \frac{1}{N} \sum_{i=1}^{N} \sum_{j=1}^{L} y_j \times
log(f(y_j|x_i)) + (1-y_j) \times log(1-f(y_j|x_i))
\end{equation}

where $N=|\mathcal{X}|$ is the total number of training examples and $f(y_j|x_i)$ is the probability that tag $y_j$ is related to SO post $x_i$. The objective captures the binary cross-entropy loss on all the training examples and can be optimized by gradient descent via back-propagation. Note that the gradient flow through the multi-label classifier as well as the PTMs used to process \textit{Title}, \textit{Description}, and \textit{Code}. The parameters of both the tag predictor after the fusion layer and the PTMs are updated during the training process of \toolnamenospace.

Given an input $x_i$, \toolname produces a vector corresponding to all the tags. An element $f(y_j|x_i)$ in the vector corresponds to the probability that tag $y_j$ is relevant with SO post $x_i$. Stack Overflow sets a limit $k$ for the number of tags a post can have, and we rank the tags in descending order according to their probabilities produced by \toolnamenospace. The top $k$ tags with the highest probabilities are recommended to a SO post.
\section{Experimental Settings}
\label{sec:experiment}

This section describes the dataset investigated in our experiment, the commonly-used evaluation metrics of a tag recommendation technique, and the implementation details of all considered models.

\subsection{Data Preparation}
\label{subsec:data_preparatin}
To ensure a fair comparison to the current state-of-the-art approach~\cite{post2vec}, we select the same dataset as Xu et al. as the benchmark.
The original data is retrieved from the snapshot of the Stack Overflow dump versioned on September 5, 2018.\footnote{\url{https://archive.org/details/stackexchange}}
A tag is regarded as \textit{rare} if its occurrence is less than a pre-defined threshold $\theta$. The intuition is that if a tag appears very infrequently in such a large corpus of posts (over 11 million posts in total), it is likely to be an incorrectly-created tag that developers do not broadly recognize. Therefore, we remove such rare tags as they may reduce the quality of the training and test data. 
Following the same procedure as \cite{post2vec, tagdc}, we set the threshold $\theta$ for deciding a rare tag as 50, and we remove all the rare tags of a post, and we remove the posts which contain rare tags only from the dataset. In the end, we have identified 29,357 rare tags and 23,687 common tags in total.
As a result, we obtained a dataset consisting of 10,379,014 posts. We select 100,000 latest posts as the test data and used the rest of 10,279,014 posts as the training data. 

\subsection{Evaluation Metrics}
\label{subsec:metrics}
Previous studies of tag recommendation~\cite{post2vec, tagdc,tagcnn} on SQA sites use $Precision@k$, $Recall@k$, $F1$-$score@k$ for evaluating the performance of the approaches. To be consistent, we follow the same evaluation metrics in this work. Formally speaking, given a corpus of SO posts, $\mathcal{X} = \{x_1,...x_n\}$, we report $Precision@k_i$, $Recall@k_i$, $F1$-$score@k_i$ on each post $x_i$ respectively where $0 \leq i \leq n$ and calculate the average of $Precision@k_i$, $Recall@k_i$, $F1$-$score@k_i$ as $Precision@k$, $Recall@k$, $F1$-$score@k$ to be the final measure.

\vspace{2mm}

\noindent\textbf{Precision@k}
measures the average ratio of predicted ground truth tags among the list of the top-k recommended tags. For the $i$th post in the test dataset, we denote its ground truth tags of a particular post by $GT_{i}$ and predicted top-k tags of the model by $Tag_{i}^{k}$. We calculate $Precision@k_i$ as: 

\vspace{1mm}
\begin{equation}
 Precision@k_i =
\frac{  \lvert  GT_{i} \cap Tag_{i}^{k} \rvert }{k}  
\end{equation}
\vspace{1mm}

\noindent Then we average all the values of $Precision@k_i$:

\vspace{1mm}
\begin{equation}
Precision@k = \frac{ \sum_{i=1 }^{\lvert \mathcal{X}\rvert  }Precision@k_i }{ \lvert \mathcal{X} \rvert }
\end{equation}
\vspace{1mm}

\noindent\textbf{Recall@k} reports the proportion of correctly predicted ground truth tags found in the list of ground truth tags. The original formula of $Recall@k_i$ has a notable drawback: the $Recall$ score would be capped to be small when the value of $k$ is smaller than the number of ground truth tags. In the past literature~\cite{post2vec,tagcnn,tagdc}, a modified version of $Recall@k$ is commonly adopted as indicated in Equations \ref{eq:recall1} and \ref{eq:recall2}. We have adopted the modified $Recall@k$ in our work, which is as same as the one used to evaluate the current state-of-the-art approach in ~\cite{post2vec}.

\vspace{1mm}
\begin{equation}
 \label{eq:recall1}
Recall@k_i =   \begin{cases}
      \frac{| GT_{i} \cap Tag_{i}^{k}}  {k}| & \text{if } |GT_{i}| > k\\
      \frac{| GT_{i} \cap Tag_{i}^{k} | }{|GT_{i}|} & \text{if } |GT_{i}| \leq k\\
    \end{cases} \\
\end{equation}
\vspace{1mm}

\begin{equation}
\label{eq:recall2}
Recall@k=  \frac{ \sum_{i=1 }^{|\mathcal{X}|}Recall@k_i}{ | \mathcal{X} |}
\end{equation}
\vspace{1mm}

\noindent\textbf{F1-score@k} is the harmonic mean of $Precision@k$ and $Recall@k$ and it is usually considered as a summary metric. It is formally defined as:

\vspace{1mm}
\begin{equation}  \label{eq:f11}
F1\text{-}score@k_i=  2 \times \frac{ Precision@k_i \times Recall@k_i}{ Precision@k_i + Recall@k_i}
\end{equation}

\vspace{1mm}
\begin{equation}  \label{eq:f12}
F1\text{-}score@k=  \frac{ \sum_{i=1 }^{|S|}F1\text{-}score@k_i}{ | \mathcal{X} |}
\end{equation}
\vspace{1mm}

A large volume of literature on tag recommendation of SQA sites evaluates the result with k equals to 5, and 10 \cite{tagdc}. However, the number of the tags for Stack Overflow post is not allowed to be greater than 5; thus, we set the maximum value of k to 5, and we evaluate $k$ on a set of values such that $k \in \{1,2,3,4,5\}$.

\subsection{Implementation}\label{sec:implementation}
To answer the three research questions mentioned in Section \ref{sec:intro}, we trained eight variants of \toolnamenospace. Details about each variant model is summarized in Table \ref{tab:variants}.

For RQ1, we trained five variants of \toolname by using different PTMs as the encoders (i.e., CodeBERT$_{ALL}$, ALBERT$_{ALL}$, BERT$_{ALL}$, RoBERTa$_{ALL}$, BERTOverflow$_{ALL}$ in Table~\ref{tab:variants}) and empirically investigate their performance.
More specifically, each variant follows the triplet architecture as illustrated in Figure \ref{fig:architecture}.

For RQ2, we compared the performance of \toolname (with the best performing PTM) with the state-of-the-art approach, namely Post2Vec. To reproduce the baseline introduced in Section~\ref{sec:post2vec}, we re-use the replication package~\footnote{\url{https://github.com/maxxbw54/Post2Vec}} released by the original authors.

After the experiments of RQ1, we found that CodeBERT$_{ALL}$ had the best performance. Thus, in RQ3 we developed three ablated models, CodeBERT$_{NoTitle}$, CodeBERT$_{NoDesp}$, and CodeBERT$_{NoCode}$, (as shown in Table~\ref{tab:variants}). Notice that since each ablated model only contains two components, they are implemented with a Twin architecture. To provide more details about the Twin architecture, it is implemented with two PTM encoders, whereas the original design of \toolname involves three PTM encoders. The rest of the design is much similar, where we concatenate the component representation obtained from each encoder and trains a multi-label classifier. 

\begin{table}[h!]
\caption{ Variants of \toolnamenospace }
    \label{tab:variants}
    \resizebox{\columnwidth}{!}{%
    \begin{tabular}{cccccc}
    \toprule
    Model Name         & BERT-Base Model    & Considered Components  & Architecture   \\
    \midrule
BERT$_{ALL}$    & BERT     & Title, Description,Code  & Triplet \\
RoBERTa$_{ALL}$    & RoBERTa     & Title, Description,Code  & Triplet \\
ALBERT$_{ALL}$    & ALBERT     & Title, Description,Code  & Triplet \\
CodeBERT$_{ALL}$    & CodeBERT     & Title, Description,Code  & Triplet \\
BERTOverflow$_{ALL}$    & BERTOverflow     & Title, Description,Code  & Triplet \\
CodeBERT$_{NoTitle}$    & CodeBERT     & Description,Code  & Twin \\
CodeBERT$_{NoDesp}$    & CodeBERT     & Title, Code  & Twin \\
CodeBERT$_{NoCode}$    & CodeBERT     & Title, Description  & Twin \\

\bottomrule       

    \end{tabular}
}
\end{table}

All the variants of \toolname are implemented with PyTorch V.1.10.0\footnote{\url{https://pytorch.org}} and HuggingFace Transformer library V.4.12.3\footnote{\url{https://huggingface.co}}. Considering the extensive amount of the data set, we only trained the models for one epoch at the fine-tuning stage. For each variant, we set the batch size as 64. We set the initial learning rate as 7E-05 and applied a linear scheduler to control the learning rate at run time.

\section{Experimental Results}
\label{sec:results}
In this section, we conduct experiments to evaluate the performance of the five variants of our proposed framework and the baseline approach. We further conduct an ablation study on our best-performing variant. We present the mean values of $Precision@k$, $Recall@k$, and $F1$-$Score@k$ for each model. Based on the results, we answer the research questions presented in Section~\ref{sec:intro}.

\subsection*{RQ1. Out of the five variants of \toolname with different PTMs, which gives the best performance?}

\noindent\textbf{Motivation}
BERT-based pre-trained language models have witnessed great success across multiple SE-related tasks. To the best of our knowledge, our work is the first that leverages PTMs in recommending tags for SQA sites. Past studies have shown that different PTMs have different strengths and weaknesses. For example, Mosel et al.~\cite{sebert} found BERTOverflow outperforms the general-purpose PTM, BERT, by a substantial margin in issue type prediction and the commit intent prediction. Mosel et al. further reported that the vocabulary of SE domain-specific PTMs (i.e., BERTOverflow) contain many programming-related vocabularies such as \textit{jvm, bugzilla and debug} which are absent in the vocabulary of BERT \cite{bert}. However, Yang et al. reported that the generic PTMs BERT performs better than domain-specific models, i.e., BERTOverflow in API review classification task \cite{chengran2022saner}. Domain-specific PTMs may understand the text from the target domain better, but general-purpose data is likely to be pre-trained on larger data. Thus, the efficacy of pre-trained models on this task remain unclear. Since the underlying BERT-based PTM of \toolname is replaceable, it evokes our interest in investigating the effectiveness of different PTMs under the \toolname architecture and finding the most suitable PTM for SO posts tag recommendation.

\begin{table}[h]
\vspace{-2mm}
\centering
\caption{Comparison of all variants of \toolname with a triplet architecture and the baseline approach Post2Vec.}
\vspace{-2mm}
\label{tab:all_results}
\begin{tabular}{c|ccccc}
\hline
\multirow{2}{*}{Model Name} & \multicolumn{5}{c}{Precision@k}                                                                                          \\ \cline{2-6} 
                            & \multicolumn{1}{c|}{P@1}   & \multicolumn{1}{c|}{P@2}   & \multicolumn{1}{c|}{P@3}   & \multicolumn{1}{c|}{P@4}   & P@5   \\ \hline
\textbf{CodeBERT$_{ALL}$}            & \multicolumn{1}{c|}{\textbf{0.848}} & \multicolumn{1}{c|}{\textbf{0.701}} & \multicolumn{1}{c|}{\textbf{0.579}} & \multicolumn{1}{c|}{\textbf{0.486}} & \textbf{0.415} \\ \hline
BERT$_{ALL}$                & \multicolumn{1}{c|}{0.845} & \multicolumn{1}{c|}{0.696} & \multicolumn{1}{c|}{0.575} & \multicolumn{1}{c|}{0.482} & 0.413 \\ \hline
RoBERTa$_{ALL}$             & \multicolumn{1}{c|}{0.843} & \multicolumn{1}{c|}{0.694} & \multicolumn{1}{c|}{0.571} & \multicolumn{1}{c|}{0.478} & 0.409 \\ \hline
BERTOverflow$_{ALL}$        & \multicolumn{1}{c|}{0.725} & \multicolumn{1}{c|}{0.592} & \multicolumn{1}{c|}{0.489} & \multicolumn{1}{c|}{0.412} & 0.354 \\ \hline
ALBERT$_{ALL}$              & \multicolumn{1}{c|}{0.748} & \multicolumn{1}{c|}{0.586} & \multicolumn{1}{c|}{0.469} & \multicolumn{1}{c|}{0.386} & 0.327 \\ \hline
Post2Vec                    & \multicolumn{1}{c|}{0.786}  & \multicolumn{1}{c|}{0.628}  & \multicolumn{1}{c|}{0.507}  & \multicolumn{1}{c|}{0.421}  & 0.359  \\ \hline \hline

\multirow{2}{*}{Model Name} & \multicolumn{5}{c}{Recall@k}                                                                                             \\ \cline{2-6} 
                            & \multicolumn{1}{c|}{R@1}   & \multicolumn{1}{c|}{R@2}   & \multicolumn{1}{c|}{R@3}   & \multicolumn{1}{c|}{R@4}   & R@5   \\ \hline
\textbf{CodeBERT$_{ALL}$}            & \multicolumn{1}{c|}{\textbf{0.848}} & \multicolumn{1}{c|}{\textbf{0.756}} & \multicolumn{1}{c|}{\textbf{0.724}} & \multicolumn{1}{c|}{\textbf{0.733}} & \textbf{0.757} \\ \hline
BERT$_{ALL}$                & \multicolumn{1}{c|}{0.845} & \multicolumn{1}{c|}{0.750} & \multicolumn{1}{c|}{0.719} & \multicolumn{1}{c|}{0.728} & 0.752 \\ \hline
RoBERTa$_{ALL}$             & \multicolumn{1}{c|}{0.843} & \multicolumn{1}{c|}{0.747} & \multicolumn{1}{c|}{0.714} & \multicolumn{1}{c|}{0.722} & 0.746 \\ \hline
BERTOverflow$_{ALL}$        & \multicolumn{1}{c|}{0.725} & \multicolumn{1}{c|}{0.635} & \multicolumn{1}{c|}{0.607} & \multicolumn{1}{c|}{0.619} & 0.644 \\ \hline
ALBERT$_{ALL}$              & \multicolumn{1}{c|}{0.748} & \multicolumn{1}{c|}{0.630} & \multicolumn{1}{c|}{0.588} & \multicolumn{1}{c|}{0.588} & 0.605 \\ \hline
Post2Vec                    & \multicolumn{1}{c|}{0.786}  & \multicolumn{1}{c|}{0.678}  & \multicolumn{1}{c|}{0.636}  & \multicolumn{1}{c|}{0.639}  & 0.659  \\ \hline \hline
\multirow{2}{*}{Model Name} & \multicolumn{5}{c}{F1-score@k}                                                                                           \\ \cline{2-6} 
                            & \multicolumn{1}{c|}{F@1}   & \multicolumn{1}{c|}{F@2}   & \multicolumn{1}{c|}{F@3}   & \multicolumn{1}{c|}{F@4}   & F@5   \\ \hline
\textbf{CodeBERT$_{ALL}$}            & \multicolumn{1}{c|}{\textbf{0.848}} & \multicolumn{1}{c|}{\textbf{0.719}} & \multicolumn{1}{c|}{\textbf{0.625}} & \multicolumn{1}{c|}{\textbf{0.561}} & \textbf{0.513} \\ \hline
BERT$_{ALL}$                & \multicolumn{1}{c|}{0.845} & \multicolumn{1}{c|}{0.714} & \multicolumn{1}{c|}{0.621} & \multicolumn{1}{c|}{0.557} & 0.510 \\ \hline
RoBERTa$_{ALL}$             & \multicolumn{1}{c|}{0.843} & \multicolumn{1}{c|}{0.711} & \multicolumn{1}{c|}{0.617} & \multicolumn{1}{c|}{0.553} & 0.505 \\ \hline
BERTOverflow$_{ALL}$        & \multicolumn{1}{c|}{0.725} & \multicolumn{1}{c|}{0.606} & \multicolumn{1}{c|}{0.527} & \multicolumn{1}{c|}{0.475} & 0.427 \\ \hline
ALBERT$_{ALL}$              & \multicolumn{1}{c|}{0.748} & \multicolumn{1}{c|}{0.600} & \multicolumn{1}{c|}{0.506} & \multicolumn{1}{c|}{0.447} & 0.406 \\ \hline
Post2Vec                    & \multicolumn{1}{c|}{0.786}  & \multicolumn{1}{c|}{0.646}  & \multicolumn{1}{c|}{0.549}  & \multicolumn{1}{c|}{0.488}  & 0.445  \\  \hline
\end{tabular}\vspace{-4mm}
\end{table}
\noindent \textbf{Results and Analysis} To answer the question, we report the performance of five variants of \toolname, which are implemented with different PTMs, which are three general-purpose PTMs (BERT, RoBERTa, and ALBERT) and two SE-specific PTMs (CodeBERT and BERTOverflow). 
These variants are implemented with the Triplet architecture, which considers Title, Description, and Code as input. We refer to these variant models as BERT$_{ALL}$, RoBERTa$_{ALL}$, ALBERT$_{ALL}$, CodeBERT$_{ALL}$, and  BERTOverflow$_{ALL}$, respectively. 

Table \ref{tab:all_results} illustrates the obtained results of all variants. We observed that leveraging CodeBERT consistently outperformed other models in all evaluation metrics. BERT$_{ALL}$ and RoBERTa$_{ALL}$ showed worse performance than CodeBERT$_{ALL}$ only by a small margin. In terms of $F1$-$score@5$, CodeBERT$_{ALL}$, BERT$_{ALL}$, and RoBERTa$_{ALL}$ achieved 0.513, 0.510 and 0.505, respectively. BERTOverflow$_{ALL}$ performed slightly better than ALBERT$_{ALL}$, and ALBERT is the worst-performing model. BERTOverflow$_{ALL}$ and ALBERT$_{ALL}$ only achieved 0.427 and 0.406 in $F1$-$score@5$, which are lower than CodeBERT, BERT and RoBERTa by a large margin.

The potential reason for why CodeBERT$_{ALL}$ achieved the best performance is that it is a bi-modal PTM for SE domain that has leveraged both natural language and programming language at the pre-training stage. Differently, the other models are trained with natural language data only. Since \toolname uses both natural and programming languages as input for tag recommendations, the nature of the input of the task well matches with CodeBERT.

ALBERT$_{ALL}$ and BERTOverflow$_{ALL}$ are largely outperformed by the rest of the variants. A potential reason could be that ALBERT includes fewer parameters since the design of ALBERT aspires to address the GPU memory limitation. BERTOverflow follows the same architecture as BERT, but BERT$_{ALL}$ performs better than BERTOverflow$_{ALL}$ by a large margin. By intuition, BERTOverflow is formulated with a much more appropriate vocabulary for the SE domain, which would also produce promising results, but the experimental results may have indicated that BERTOverflow still required more training. It is potentially caused by the quality and the size of the datasets used at the pre-training stage. BERTOverflow is pre-trained with 152 million sentences from SO. BERT is trained on the entire English Wikipedia and the Book Corpus dataset, written by professionals and constantly reviewed. However, sentences from SO could be written by arbitrary authors and the existence of in-line code within a post would introduce extra noise. Additionally, the training corpus of BERT contains 3.3 billion words in total, and the average sentence length of BookCorpus is 11 words. By estimation, the training corpus of BERT is likely to be twice more than BERTOverflow.

Although the CodeBERT and BERTOverflow both are SE domain-specific PTMs, fundamentally,  are different from each other. In architecture-wise, CodeBERT is initialized with RoBERTa's parameters and adds a second phase of pre-training on the CodeSearchNet dataset \cite{codesearchnet}, whereas BERTOverflow is trained from scratch. Moreover, both BERT and RoBERTa are models designed for natural language only. Although we include code snippets as input, BERT and RoBERTa still achieve outstanding results (slightly worse than CodeBERT). It possibly indicates that pre-trained models trained with a large scale of natural language data also be helpful for programming language modeling. 
\vspace{-4mm}
\begin{tcolorbox}
 \textbf{Answers to RQ1}: Among the five considered PTMs of \toolname, the one implemented with CodeBERT produces the best performance.
\end{tcolorbox}
\vspace{-1.8em}

\subsection*{RQ2. How is the performance of \toolname compared to the state-of-the-art approach in Stack Overflow tag recommendation?}

\noindent\textbf{Motivation} The current state-of-the-art approach for recommending tags of SO posts is implemented based on Convolutional Neural Network and trained from scratch~\cite{post2vec}. However, Transformer-based PTMs are strengthened with the self-attention mechanism and transferred pre-train knowledge. In this research question, we investigate whether the variants of \toolname can achieve better performance than the current state-of-the-art approach.

\noindent\textbf{Results and Analysis}
As presented in Table~\ref{tab:all_results}, the best performing variant of \toolnamenospace, i.e., CodeBERT$_{ALL}$, substantially outperformed Post2Vec in terms of $F1$-$score@k$ (k from 1 to 5) by 7.9\%, 11.3\%, 13.8\%, 15.0\% and 15.3\%, respectively. Furthermore, BERT$_{ALL}$ and RoBERTa$_{ALL}$ also surpassed Post2Vec. BERT$_{ALL}$ improved the $F1$-$score@1$-$5$ of Post2Vec by 7.5\%, 10.5\%, 13.1\%, 14.1\% and 14.6\%; and RoBERTa$_{ALL}$ improved by 7.3\%, 10.1\%, 12.4\%, 13.3\% and 13.5\%; However, we also observe that not all PTMs are able to achieve outstanding performance under \toolnamenospace. Post2Vec outperformed BERTOverflow$_{ALL}$ by 8.4\%, 6.6\%, 8.5\%, 2.7\%, 4.7\% and outperformed ALBERT$_{ALL}$ by 5.1\%, 7.7\%, 8.5\%, 9.2\%, 9.6\% in $F1$-$score@1$--$5$.


Different from Post2Vec, \toolname involves a vast amount of knowledge accumulated from the dataset used for pre-training. \toolname leverages PTMs to extract feature vectors and optimize the post representation during the fine-tuning stage, whereas Post2Vec learns post representations from scratch. Thus, PTMs provide a more decent initial setting. Furthermore, CodeBERT provides in-domain knowledge of SE. Our results indicate that the knowledge learned in the pre-training stage is valuable to the success of the tag recommendation task. 

Another potential reason for the superior performance of \toolname is that transformer-based models are more powerful than CNN in capturing long-range dependencies~\cite{transformer}. The architecture of BERT-based PTMs is inherited from a Transformer. One of the critical concepts of Transformers is the \textit{self-attention mechanism}, which enables its ability to capture long dependencies among all input sequences.
Our results demonstrate the effectiveness and generalizability of transfer learning and reveal that the PTMs can achieve outstanding performance in the tag recommendation task for SO posts.

\vspace{-3mm}
\begin{tcolorbox}
    \textbf{Answers to RQ2}: 
    CodeBERT$_{ALL}$, BERT$_{ALL}$ and RoBERTa$_{ALL}$ outperform the state-of-the-art approach by a substantial margin. However, BERTOverflow$_{ALL}$ and ALBERT$_{ALL}$ demonstrated worse performance than the state-of-the-art approach.
\end{tcolorbox}
\vspace{-5mm}

\subsection*{RQ3. Which component of post benefits \toolname the most?}

\textbf{Motivation} 
\toolname is designed with a triplet architecture where each component of a post, i.e., Title, Description, and Code, are modeled by utilizing separate PTMs. 
Title, Description, and Code snippets complement each other and describe the post from its own perspective. Title summarizes the question with a few words; Description further expands the content from the Title; Code snippets often is a real example of the problem. Thus it motivates us to investigate which component produces the most critical contribution in the \toolname framework.

\begin{table}[]
\centering
\caption{Experiment Results of RQ3, the ablation study conducted for each post component.
}
\label{tab:rq3}
\vspace{-4mm}
\begin{tabular}{c|ccccc}
\hline
\multirow{2}{*}{Model Name} & \multicolumn{5}{c}{Precision@k}                                                                            \\ \cline{2-6} 
                            & \multicolumn{1}{c|}{P@1}   & \multicolumn{1}{c|}{P@2}   & \multicolumn{1}{c|}{P@3}   & \multicolumn{1}{c|}{P@4}   & P@5   \\ \hline
\textbf{CodeBERT$_{ALL}$}            & \multicolumn{1}{c|}{\textbf{0.848}} & \multicolumn{1}{c|}{\textbf{0.701}} & \multicolumn{1}{c|}{\textbf{0.579}} & \multicolumn{1}{c|}{\textbf{0.486}} & \textbf{0.415} \\ \hline
CodeBERT$_{NoCode}$         & \multicolumn{1}{c|}{0.823} & \multicolumn{1}{c|}{0.682} & \multicolumn{1}{c|}{0.562} & \multicolumn{1}{c|}{0.472} & 0.408 \\ \hline
CodeBERT$_{NoDesp}$         & \multicolumn{1}{c|}{0.822} & \multicolumn{1}{c|}{0.671} & \multicolumn{1}{c|}{0.549} & \multicolumn{1}{c|}{0.458} & 0.390 \\ \hline
CodeBERT$_{NoTitle}$        & \multicolumn{1}{c|}{0.808} & \multicolumn{1}{c|}{0.664} & \multicolumn{1}{c|}{0.547} & \multicolumn{1}{c|}{0.460} & 0.394 \\ \hline \hline
\multirow{2}{*}{Model Name} & \multicolumn{5}{c}{Recall@k}                                                                                             \\ \cline{2-6} 
                            & \multicolumn{1}{c|}{R@1}   & \multicolumn{1}{c|}{R@2}   & \multicolumn{1}{c|}{R@3}   & \multicolumn{1}{c|}{R@4}   & R@5   \\ \hline
\textbf{CodeBERT$_{ALL}$}            & \multicolumn{1}{c|}{\textbf{0.848}} & \multicolumn{1}{c|}{\textbf{0.756}} & \multicolumn{1}{c|}{\textbf{0.724}} & \multicolumn{1}{c|}{\textbf{0.733}} & \textbf{0.757} \\ \hline
CodeBERT$_{NoCode}$         & \multicolumn{1}{c|}{0.823} & \multicolumn{1}{c|}{0.733} & \multicolumn{1}{c|}{0.702} & \multicolumn{1}{c|}{0.712} & 0.737 \\ \hline
CodeBERT$_{NoDesp}$         & \multicolumn{1}{c|}{0.822} & \multicolumn{1}{c|}{0.723} & \multicolumn{1}{c|}{0.686} & \multicolumn{1}{c|}{0.693} & 0.714 \\ \hline
CodeBERT$_{NoTitle}$        & \multicolumn{1}{c|}{0.808} & \multicolumn{1}{c|}{0.715} & \multicolumn{1}{c|}{0.683} & \multicolumn{1}{c|}{0.693} & 0.718 \\ \hline \hline
\multirow{2}{*}{Model Name} & \multicolumn{5}{c}{F1-score@k}                                                                                           \\ \cline{2-6} 
                            & \multicolumn{1}{c|}{F@1}   & \multicolumn{1}{c|}{F@2}   & \multicolumn{1}{c|}{F@3}   & \multicolumn{1}{c|}{F@4}   & F@5   \\ \hline
\textbf{CodeBERT$_{ALL}$}            & \multicolumn{1}{c|}{\textbf{0.848}} & \multicolumn{1}{c|}{\textbf{0.719}} & \multicolumn{1}{c|}{\textbf{0.625}} & \multicolumn{1}{c|}{\textbf{0.561}} & \textbf{0.513} \\ \hline
CodeBERT$_{NoCode}$         & \multicolumn{1}{c|}{0.823} & \multicolumn{1}{c|}{0.699} & \multicolumn{1}{c|}{0.607} & \multicolumn{1}{c|}{0.545} & 0.500 \\ \hline
CodeBERT$_{NoDesp}$         & \multicolumn{1}{c|}{0.822} & \multicolumn{1}{c|}{0.688} & \multicolumn{1}{c|}{0.593} & \multicolumn{1}{c|}{0.530} & 0.483 \\ \hline
CodeBERT$_{NoTitle}$        & \multicolumn{1}{c|}{0.808} & \multicolumn{1}{c|}{0.680} & \multicolumn{1}{c|}{0.591} & \multicolumn{1}{c|}{0.531} & 0.487 \\ \hline
\end{tabular}\vspace{-3mm}
\end{table}

\noindent \textbf{Results and Analysis} To answer this research question, we conduct an ablation study to investigate the importance of each component, i.e., Title, Description, and Code, respectively. Note that \toolname is implemented with a triplet architecture by default. To answer this research question, we modified it to a twin architecture to fit two considered components at a time. We train three ablated models with our identified best-performing PTM, i.e., CodeBERT.

\begin{figure}[]
    \label{fig:graph}
	\includegraphics[width=1\linewidth]{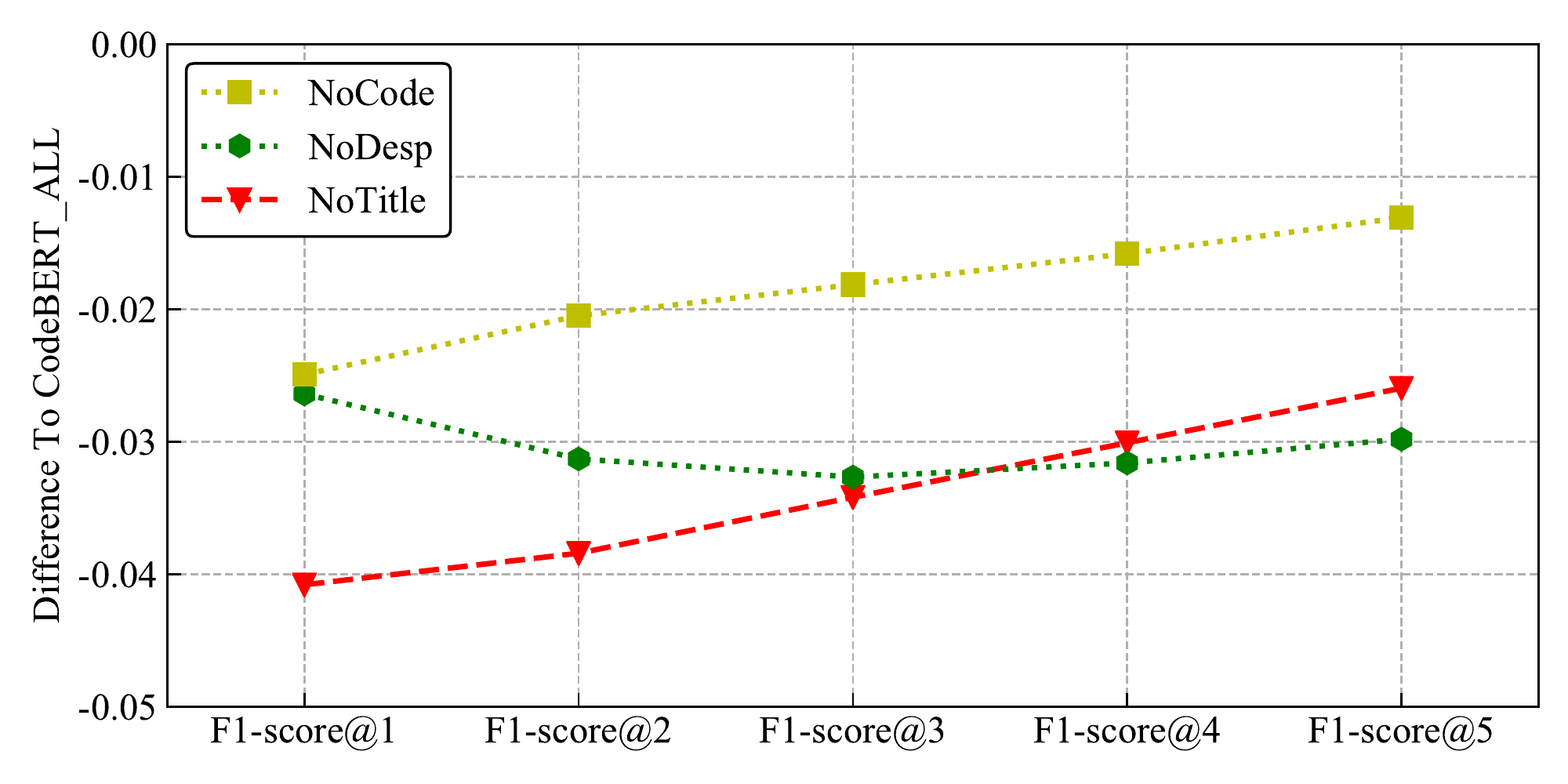}
	\vspace{-3mm}
	\caption{A line chart demonstrate performance difference in $F1$-$score@k$ between each ablated models and CodeBERT$_{ALL}$, where $k \in \{1,2,3,4,5\}$. The value on the y axis is calculated using the corresponding score of the candidate ablated model minus the corresponding score of CodeBERT$_{ALL}$ .}
	\label{fig:f1}
	\vspace{-5mm}
\end{figure}

The results for RQ3 are presented in Table \ref{tab:rq3}. From the table, we identified that CodeBERT$_{ALL}$ remained the best performing model on all the evaluation metrics. To provide a more intuitive understanding of the result, we further illustrate the performance gap of $F1$-$score@1$--$5$ between the ablated models and CodeBERT$_{ALL}$ by visualizing in Figure \ref{fig:f1}, where the value on the y axis is calculated using the score of CodeBERT$_{ALL}$ minus the score of the ablated model. CodeBERT$_{NoCode}$ yielded the most promising performance among the ablated models, which implies that the code snippets are beneficial, but they are the least significant among all three components.

An interesting finding is that CodeBERT$_{NoDesc}$ performed better in $F1$-$score@1$--$3$ and CodeBERT$_{NoTitle}$ performed better in $F1$-$score@4$--$5$, which suggests that Title is more important for boosting the performance of $F1$-$score@1$--$3$ and Description is more critical for improving $F1$-$score@4$--$5$. 
A possible explanation could be that Title always succinctly describes a post's central message, which could directly help the system predict the top few tags. Description is usually much longer and elaborates the Title with more explanations; thus, it is more beneficial to recommend the last few tags. Moreover, as CodeBERT$_{ALL}$ is the best performing model, which suggests that Code is beneficial in the tag recommendation task of SO.

\vspace{-1mm}
\begin{tcolorbox}
    \textbf{Answers to RQ3}: 
    Under \toolnamenospace, Title and Description are more important than Code for tag recommendation. Title plays the most important role in predicting the top 3 most relevant tags, and the contribution of Description increases when the number of predicted tags increases from 4. Still, considering all the components can achieve the best performance.
\end{tcolorbox}

\vspace{-2mm}
\section{Discussion}
\label{sec:discussion}

\subsection{Error Analysis}
We conduct qualitative analysis to illustrate the capability of our proposed framework \toolnamenospace. Take a Stack Overflow post\footnote{\url{https://stackoverflow.com/questions/51910978}} titled \emph{Pass Input Function to Multiple Class Decorators} in the test dataset as an example.
The ground truth tags of the post are {\tt decorator}, {\tt memoization}, {\tt python}, {\tt python-2.7}, and {\tt python-decorators}. The tags predicted by Post2Vec are {\tt class}, {\tt timer}, {\tt python-decorators}, {\tt decorator}, and {\tt  python}
 while \toolname gives the exact prediction as to the ground truth tags. Although the word \emph{memoization} has occurred several times in the post, Post2Vec still failed to capture its existence and the connection between \emph{memoization} and \emph{decorator}. Moreover, we found that the CodeSearchNet database \cite{codesearchnet} which is used to pre-train CodeBERT includes source code files that relate to both \emph{memoization} and \emph{decorator} \footnote{\url{https://github.com/nerdvegas/rez/blob/1d3b846d53b5b5404edfe8ddb9083f9ceec8c5e7/src/rez/utils/memcached.py\#L248-L375}}, This potentially could indicate that the pre-trained knowledge learned by CodeBERT is beneficial for our task.

\subsection{\textbf{Threats to Validity}}

\subsubsection*{\textbf{Threats to internal validity}} To ensure we implement the baseline (i.e., Post2Vec) correctly, we reused the official replication package released by the Xu et al.\footnote{\url{https://github.com/maxxbw54/Post2Vec}} To instantiate variants of \toolname with different pre-trained models, we utilized a widely-used deep learning library \textit{Hugging Face}.\footnote{\url{https://huggingface.co/}} Similar to prior studies~\cite{wang2018entagrec,wang2015tagcombine,post2vec}, our work assumes that the tags are labeled correctly by users in Stack Overflow. However, some tags are potentially mislabelled. Still, we believe that Stack Overflow’s effective crowdsourcing process helps to reduce the number of such cases, and we further minimize this threat by discarding rare tags and posts (as described in Section~\ref{subsec:data_preparatin}).

\subsubsection*{\textbf{Threats to external validity}} 
We analyzed Stack Overflow, the largest SQA site, with a massive amount of questions. These questions cover diverse discussions on various software topics. As software technologies evolve fast, our results may not generalize to those newly emerging topics. Instead, our framework can adapt to new posts by fine-tuning models on more and new questions.

\subsubsection*{\textbf{Threats to construct validity}} Threats to construct validity are related to the suitability of our evaluation metrics. $Precision@k$, $Recall@k$, and $F1$-$score@k$ are widely used to evaluate many tag recommendation approaches in software engineering~\cite{xia2013tag,wang2018entagrec,wang2015tagcombine,post2vec}. Thus, we believe the threat is minimal.

\subsection{Lessons Learned}
\label{subsec:lessons}

\subsubsection*{\textbf{Lesson \#1}}
\textbf{\textit{Pre-trained language models are effective in tagging SO posts.}}
The tag recommendation task for SQA sites has been extensively studied in the last decade~\cite{tagdc, tagcnn, tagcombine, post2vec}. Researchers have tackled the problem via a range of techniques, e.g., collaborative filtering~\cite{tagdc} and deep learning~\cite{post2vec}. Furthermore, these techniques usually involve separate training of each component. Our experiment results have demonstrated that simply fine-tuning the pre-trained Transformer-based model can achieve state-of-the-art performance, even if there are thousands of tags. CodeBERT, BERT, and RoBERTa are capable of providing promising results for tag recommendation. Even though BERT and RoBERTa did not leverage programming language at the pre-training stage. 

Although PTMs are already widely adopted in SE tasks, most tasks are formulated as either {\em binary} classification problems or {\em multi-class} classification problems. In binary or multi-class classification problems the label classes are mutually exclusive, whereas for multi-label classification problems each data point may belongs to several labels simultaneously. We encourage practitioners to leverage pre-trained models in the {\em multi-label} classification settings where the size of the label set could go beyond thousands. Moreover, our experiments also validate the generalizability of pre-trained models. We recommend practitioners apply pre-trained models in more SE tasks and consider fine-tuning pre-trained models as one of their baselines. 

\subsubsection*{\textbf{Lesson \#2}}
\textbf{\textit{CodeBERT is recommended as one of the first attempts for SQA-related tasks which involve both natural language and programming language.}} CodeBERT is trained on both bimodal and unimodal data. Our results showed that CodeBERT consistently outperforms other pre-trained models as well as the recent state-of-the-art deep learning model in tag recommendation task of Stack Overflow. We found that SE domain-specific model, i.e., CodeBERT is powerful, and researchers are recommended to use CodeBERT as one of their first attempts when their tasks are based on SQA data and involve both NL and PL.

\subsubsection*{\textbf{Lesson \#3}}
\textbf{\textit{All components of a post from Stack Overflow are valuable pieces of semantics.}} Most previous literature has removed the code snippets from the pre-training process because they are considered noisy, poorly structured, and written in many different programming languages~\cite{tagcnn, tagdc, tagcombine, tagmulrec}. However, our results show that code snippets are also beneficial to capture the semantics of SO posts and further boost the performance of the tag recommendation task. We encourage researchers to consider both the natural and programming languages parts of a post when analyzing SQA sites.

\vspace{-2mm}
\section{related work}

\label{sec:related_work}
\subsection{Pre-trained Models in SE}
Recent works~\cite{biobert,scibert,clinicalbert} have shown that the in-domain knowledge acquired by PTMs is valuable in improving the performance on domain-specific tasks, such as ClinicalBERT~\cite{clinicalbert} for clinical text, SciBERT~\cite{scibert} for scientific text, and BioBERT \cite{biobert} for biomedical text. Evoked by the success of PTMs in other domains, researchers start the work in the SE domain-specific PTMs~\cite{CodeBERT, sebert, cbert, cubert, bertoverflow}. 
Developers are free to create and pick arbitrary identifiers, making texts in the SE field frequently involve technical jargon and tokens from the programming languages~\cite{shi2022identifier}. This makes models pre-trained on general texts (e.g., BERT~\cite{bert}) unsuitable to represent texts in the software engineering domain. Tabassum et al.~\cite{bertoverflow} provided BERTOverflow for completing tasks related to Stack Overflow. BERTOverflow~\cite{bertoverflow} inherits the architecture from BERT~\cite{bert} and is trained on more than 152 million programming-related questions extracted from Stack Overflow. Mosel et al.~\cite{sebert} aims to provide a better SE domain-specific pre-trained model than BERTOverflow~\cite{bertoverflow} and propose seBERT~\cite{sebert}. Since seBERT is developed upon the BERT$_{LARGE}$ architecture, we did not investigate the effectiveness of seBERT under the \toolname framework due to restraints on computational resources and GPU memory consumption. 

The aforementioned PTMs focus on modeling the natural language texts in the SE domain. In addition, there is a proliferation of studies on model programming languages with PTMs. 
Svyatkovskiy et al. trained GPT-C~\cite{gpt-c}, a multi-layer generative transformer model, and leveraged GPT-C to build a code completion tool. GPT-C is pre-trained on 1.2 billion lines of source code written by multiple programming languages, including Python, JavaScript, C-sharp, etc.
Feng et al. proposed CodeBERT~\cite{CodeBERT}, a bi-model pre-trained model for both natural and programming languages (NL and PL). CodeBERT is trained on bi-modal data of NL-PL pairs and a vast amount of uni-modal data. CodeBERT has two pre-training tasks: Masked Language Modeling (MLM) and Replacement Token Detection (RTD). The experimental results show that CodeBERT achieves the state-of-the-art performance on natural language code search and code document generation, and in a zero-shot setting, CodeBERT persistently outperforms RoBERTa~\cite{roberta}. 

\subsection{Tag Recommendation for SQA Sites}
Researchers have already extensively studied the tag recommendation task in the SE domain and proposed a number of approaches.
Wang et al. introduced EnTagRec~\cite{entagrec} that utilizes Bayesian inference and an enhanced frequentist inference technique. Results show that it outperformed TagCombine by a significant margin. Wang et al. then further extends EnTagRec~\cite{entagrec} to EnTagRec++~\cite{entagresplusplus}, the latter of which additionally considers user information and an initial set of tags provided by a user. Zhou et al. proposed TagMulRec~\cite{tagmulrec}, a collaborative filtering method that suggests new tags of a post based on the results of semantically similar posts. 
Li et al. proposed TagDC~\cite{tagdc}, which is implemented with two parts: TagDC-DL that leverages a content-based approach to learn a multi-label classifier with a CNN Capsule network, and TagDC-CF that utilizes collaborative filtering to focus on the tags of similar historical posts. 
Post2Vec~\cite{post2vec} distributively represents SO posts and is shown useful in tackling numerous Stack Overflow-related downstream tasks. We select Post2Vec~\cite{post2vec} as the baseline in our experiments as it achieves the state-of-the-art performance in the tag recommendation task for SO posts.
\section{Conclusion and Future Work}
\label{sec:conclusion}

In this work, we modeled the \textit{tag recommendation task of SO} as a \textit{multi-label classification problem} and introduced \toolnamenospace, a pre-trained model-based framework for tag recommendation.
We implemented five variants of \toolname with different PTMs as the underlying Encoder. Our experiment results showed that the best variant of \toolname is CodeBERT-based, and it outperforms the state-of-the-art approach by a large margin in terms of $F1$-$score@5$. Leveraging BERT and RoBERTa could also achieve desirable performance that only slightly loses CodeBERT and beat the state-of-the-art approach by a large margin. However, \toolname models implemented with BERTOverflow and ALBERT did not give promising results. Even though the BERT-based PTMs are shown to be powerful and effective, PTMs behave differently, and the selection of PTMs needs to be carefully decided.
Besides, we conducted an ablation study to investigate the contribution of each component under \toolnamenospace. The result shows that Title is the most vital part for recommending the most relevant tags, and considering all components simultaneously and separately can produce the best performance. In the future, we plan to consider more diverse information of a post into account, such as the attached pictures, author information, etc. Also, we are interested in applying \toolname on more SQA sites such as AskUbuntu\footnote{\url{https://askubuntu.com/}}, etc., to further evaluate its effectiveness and generalizability. We release our replication package\footnote{\url{https://github.com/soarsmu/PTM4Tag}} to facilitate future research.

\begin{acks}
This research / project is supported by the National Research Foundation, Singapore, under its Industry Alignment Fund – Pre-positioning (IAF-PP) Funding Initiative. Any opinions, findings and conclusions or recommendations expressed in this material are those of the authors and do not reflect the views of National Research Foundation, Singapore.
\end{acks}

\balance
\bibliographystyle{ACM-Reference-Format}
\bibliography{reference}


\begin{thebibliography}{43}


\ifx \showCODEN    \undefined \def \showCODEN     #1{\unskip}     \fi
\ifx \showDOI      \undefined \def \showDOI       #1{#1}\fi
\ifx \showISBNx    \undefined \def \showISBNx     #1{\unskip}     \fi
\ifx \showISBNxiii \undefined \def \showISBNxiii  #1{\unskip}     \fi
\ifx \showISSN     \undefined \def \showISSN      #1{\unskip}     \fi
\ifx \showLCCN     \undefined \def \showLCCN      #1{\unskip}     \fi
\ifx \shownote     \undefined \def \shownote      #1{#1}          \fi
\ifx \showarticletitle \undefined \def \showarticletitle #1{#1}   \fi
\ifx \showURL      \undefined \def \showURL       {\relax}        \fi
\providecommand\bibfield[2]{#2}
\providecommand\bibinfo[2]{#2}
\providecommand\natexlab[1]{#1}
\providecommand\showeprint[2][]{arXiv:#2}

\bibitem[\protect\citeauthoryear{Barua, Thomas, and Hassan}{Barua
  et~al\mbox{.}}{2014}]%
        {barua2014developers}
\bibfield{author}{\bibinfo{person}{Anton Barua}, \bibinfo{person}{Stephen~W
  Thomas}, {and} \bibinfo{person}{Ahmed~E Hassan}.}
  \bibinfo{year}{2014}\natexlab{}.
\newblock \showarticletitle{What are developers talking about? an analysis of
  topics and trends in stack overflow}.
\newblock \bibinfo{journal}{\emph{Empirical Software Engineering}}
  \bibinfo{volume}{19}, \bibinfo{number}{3} (\bibinfo{year}{2014}),
  \bibinfo{pages}{619--654}.
\newblock


\bibitem[\protect\citeauthoryear{Beltagy, Lo, and Cohan}{Beltagy
  et~al\mbox{.}}{2019}]%
        {scibert}
\bibfield{author}{\bibinfo{person}{Iz Beltagy}, \bibinfo{person}{Kyle Lo},
  {and} \bibinfo{person}{Arman Cohan}.} \bibinfo{year}{2019}\natexlab{}.
\newblock \bibinfo{title}{SciBERT: A Pretrained Language Model for Scientific
  Text}.
\newblock
\newblock
\showeprint[arxiv]{1903.10676}~[cs.CL]


\bibitem[\protect\citeauthoryear{Bowman, Angeli, Potts, and Manning}{Bowman
  et~al\mbox{.}}{2015}]%
        {bowman-etal-2015-large}
\bibfield{author}{\bibinfo{person}{Samuel~R. Bowman}, \bibinfo{person}{Gabor
  Angeli}, \bibinfo{person}{Christopher Potts}, {and}
  \bibinfo{person}{Christopher~D. Manning}.} \bibinfo{year}{2015}\natexlab{}.
\newblock \showarticletitle{A large annotated corpus for learning natural
  language inference}. In \bibinfo{booktitle}{\emph{Proceedings of the 2015
  Conference on Empirical Methods in Natural Language Processing}}.
  \bibinfo{publisher}{Association for Computational Linguistics},
  \bibinfo{address}{Lisbon, Portugal}, \bibinfo{pages}{632--642}.
\newblock
\urldef\tempurl%
\url{https://doi.org/10.18653/v1/D15-1075}
\showDOI{\tempurl}


\bibitem[\protect\citeauthoryear{Buratti, Pujar, Bornea, McCarley, Zheng,
  Rossiello, Morari, Laredo, Thost, Zhuang, et~al\mbox{.}}{Buratti
  et~al\mbox{.}}{2020}]%
        {cbert}
\bibfield{author}{\bibinfo{person}{Luca Buratti}, \bibinfo{person}{Saurabh
  Pujar}, \bibinfo{person}{Mihaela Bornea}, \bibinfo{person}{Scott McCarley},
  \bibinfo{person}{Yunhui Zheng}, \bibinfo{person}{Gaetano Rossiello},
  \bibinfo{person}{Alessandro Morari}, \bibinfo{person}{Jim Laredo},
  \bibinfo{person}{Veronika Thost}, \bibinfo{person}{Yufan Zhuang},
  {et~al\mbox{.}}} \bibinfo{year}{2020}\natexlab{}.
\newblock \showarticletitle{Exploring software naturalness through neural
  language models}.
\newblock \bibinfo{journal}{\emph{arXiv preprint arXiv:2006.12641}}
  (\bibinfo{year}{2020}).
\newblock


\bibitem[\protect\citeauthoryear{Devlin, Chang, Lee, and Toutanova}{Devlin
  et~al\mbox{.}}{2018}]%
        {bert}
\bibfield{author}{\bibinfo{person}{Jacob Devlin}, \bibinfo{person}{Ming{-}Wei
  Chang}, \bibinfo{person}{Kenton Lee}, {and} \bibinfo{person}{Kristina
  Toutanova}.} \bibinfo{year}{2018}\natexlab{}.
\newblock \showarticletitle{{BERT:} Pre-training of Deep Bidirectional
  Transformers for Language Understanding}.
\newblock \bibinfo{journal}{\emph{CoRR}}  \bibinfo{volume}{abs/1810.04805}
  (\bibinfo{year}{2018}).
\newblock
\showeprint[arXiv]{1810.04805}
\urldef\tempurl%
\url{http://arxiv.org/abs/1810.04805}
\showURL{%
\tempurl}


\bibitem[\protect\citeauthoryear{Feng, Guo, Tang, Duan, Feng, Gong, Shou, Qin,
  Liu, Jiang, and Zhou}{Feng et~al\mbox{.}}{2020}]%
        {CodeBERT}
\bibfield{author}{\bibinfo{person}{Zhangyin Feng}, \bibinfo{person}{Daya Guo},
  \bibinfo{person}{Duyu Tang}, \bibinfo{person}{Nan Duan},
  \bibinfo{person}{Xiaocheng Feng}, \bibinfo{person}{Ming Gong},
  \bibinfo{person}{Linjun Shou}, \bibinfo{person}{Bing Qin},
  \bibinfo{person}{Ting Liu}, \bibinfo{person}{Daxin Jiang}, {and}
  \bibinfo{person}{Ming Zhou}.} \bibinfo{year}{2020}\natexlab{}.
\newblock \showarticletitle{{C}ode{BERT}: A Pre-Trained Model for Programming
  and Natural Languages}. In \bibinfo{booktitle}{\emph{Findings of the
  Association for Computational Linguistics: EMNLP 2020}}.
  \bibinfo{publisher}{Association for Computational Linguistics},
  \bibinfo{address}{Online}, \bibinfo{pages}{1536--1547}.
\newblock
\urldef\tempurl%
\url{https://doi.org/10.18653/v1/2020.findings-emnlp.139}
\showDOI{\tempurl}


\bibitem[\protect\citeauthoryear{Huang, Tang, Shou, Gong, Xu, Jiang, Zhou, and
  Duan}{Huang et~al\mbox{.}}{2021}]%
        {CoSQA}
\bibfield{author}{\bibinfo{person}{Junjie Huang}, \bibinfo{person}{Duyu Tang},
  \bibinfo{person}{Linjun Shou}, \bibinfo{person}{Ming Gong},
  \bibinfo{person}{Ke Xu}, \bibinfo{person}{Daxin Jiang}, \bibinfo{person}{Ming
  Zhou}, {and} \bibinfo{person}{Nan Duan}.} \bibinfo{year}{2021}\natexlab{}.
\newblock \showarticletitle{CoSQA: 20, 000+ Web Queries for Code Search and
  Question Answering}. In \bibinfo{booktitle}{\emph{Proceedings of the 59th
  Annual Meeting of the Association for Computational Linguistics and the 11th
  International Joint Conference on Natural Language Processing, {ACL/IJCNLP}
  2021, (Volume 1: Long Papers), Virtual Event, August 1-6, 2021}},
  \bibfield{editor}{\bibinfo{person}{Chengqing Zong}, \bibinfo{person}{Fei
  Xia}, \bibinfo{person}{Wenjie Li}, {and} \bibinfo{person}{Roberto Navigli}}
  (Eds.). \bibinfo{publisher}{Association for Computational Linguistics},
  \bibinfo{pages}{5690--5700}.
\newblock
\urldef\tempurl%
\url{https://doi.org/10.18653/v1/2021.acl-long.442}
\showDOI{\tempurl}


\bibitem[\protect\citeauthoryear{Huang, Altosaar, and Ranganath}{Huang
  et~al\mbox{.}}{2020}]%
        {clinicalbert}
\bibfield{author}{\bibinfo{person}{Kexin Huang}, \bibinfo{person}{Jaan
  Altosaar}, {and} \bibinfo{person}{Rajesh Ranganath}.}
  \bibinfo{year}{2020}\natexlab{}.
\newblock \bibinfo{title}{ClinicalBERT: Modeling Clinical Notes and Predicting
  Hospital Readmission}.
\newblock
\newblock
\showeprint[arxiv]{1904.05342}~[cs.CL]


\bibitem[\protect\citeauthoryear{Husain, Wu, Gazit, Allamanis, and
  Brockschmidt}{Husain et~al\mbox{.}}{2020}]%
        {codesearchnet}
\bibfield{author}{\bibinfo{person}{Hamel Husain}, \bibinfo{person}{Ho-Hsiang
  Wu}, \bibinfo{person}{Tiferet Gazit}, \bibinfo{person}{Miltiadis Allamanis},
  {and} \bibinfo{person}{Marc Brockschmidt}.} \bibinfo{year}{2020}\natexlab{}.
\newblock \bibinfo{title}{CodeSearchNet Challenge: Evaluating the State of
  Semantic Code Search}.
\newblock
\newblock
\showeprint[arxiv]{1909.09436}~[cs.LG]


\bibitem[\protect\citeauthoryear{Jin, Jin, Zhou, and Szolovits}{Jin
  et~al\mbox{.}}{2020}]%
        {jin2020bert}
\bibfield{author}{\bibinfo{person}{Di Jin}, \bibinfo{person}{Zhijing Jin},
  \bibinfo{person}{Joey~Tianyi Zhou}, {and} \bibinfo{person}{Peter Szolovits}.}
  \bibinfo{year}{2020}\natexlab{}.
\newblock \showarticletitle{Is bert really robust? a strong baseline for
  natural language attack on text classification and entailment}. In
  \bibinfo{booktitle}{\emph{Proceedings of the AAAI conference on artificial
  intelligence}}, Vol.~\bibinfo{volume}{34}. \bibinfo{pages}{8018--8025}.
\newblock


\bibitem[\protect\citeauthoryear{Kanade, Maniatis, Balakrishnan, and
  Shi}{Kanade et~al\mbox{.}}{2020}]%
        {cubert}
\bibfield{author}{\bibinfo{person}{Aditya Kanade}, \bibinfo{person}{Petros
  Maniatis}, \bibinfo{person}{Gogul Balakrishnan}, {and}
  \bibinfo{person}{Kensen Shi}.} \bibinfo{year}{2020}\natexlab{}.
\newblock \showarticletitle{Learning and Evaluating Contextual Embedding of
  Source Code}. In \bibinfo{booktitle}{\emph{Proceedings of the 37th
  International Conference on Machine Learning}}
  \emph{(\bibinfo{series}{Proceedings of Machine Learning Research},
  Vol.~\bibinfo{volume}{119})}, \bibfield{editor}{\bibinfo{person}{Hal~Daumé
  III} {and} \bibinfo{person}{Aarti Singh}} (Eds.). \bibinfo{publisher}{PMLR},
  \bibinfo{pages}{5110--5121}.
\newblock
\urldef\tempurl%
\url{https://proceedings.mlr.press/v119/kanade20a.html}
\showURL{%
\tempurl}


\bibitem[\protect\citeauthoryear{Lan, Chen, Goodman, Gimpel, Sharma, and
  Soricut}{Lan et~al\mbox{.}}{2020}]%
        {albert}
\bibfield{author}{\bibinfo{person}{Zhenzhong Lan}, \bibinfo{person}{Mingda
  Chen}, \bibinfo{person}{Sebastian Goodman}, \bibinfo{person}{Kevin Gimpel},
  \bibinfo{person}{Piyush Sharma}, {and} \bibinfo{person}{Radu Soricut}.}
  \bibinfo{year}{2020}\natexlab{}.
\newblock \bibinfo{title}{ALBERT: A Lite BERT for Self-supervised Learning of
  Language Representations}.
\newblock
\newblock
\showeprint[arxiv]{1909.11942}~[cs.CL]


\bibitem[\protect\citeauthoryear{Lee, Yoon, Kim, Kim, Kim, So, and Kang}{Lee
  et~al\mbox{.}}{2019}]%
        {biobert}
\bibfield{author}{\bibinfo{person}{Jinhyuk Lee}, \bibinfo{person}{Wonjin Yoon},
  \bibinfo{person}{Sungdong Kim}, \bibinfo{person}{Donghyeon Kim},
  \bibinfo{person}{Sunkyu Kim}, \bibinfo{person}{Chan~Ho So}, {and}
  \bibinfo{person}{Jaewoo Kang}.} \bibinfo{year}{2019}\natexlab{}.
\newblock \showarticletitle{BioBERT: a pre-trained biomedical language
  representation model for biomedical text mining}.
\newblock \bibinfo{journal}{\emph{Bioinformatics}} (\bibinfo{date}{Sep}
  \bibinfo{year}{2019}).
\newblock
\showISSN{1460-2059}
\urldef\tempurl%
\url{https://doi.org/10.1093/bioinformatics/btz682}
\showDOI{\tempurl}


\bibitem[\protect\citeauthoryear{Li, Xu, Yan, and Lei}{Li
  et~al\mbox{.}}{2020}]%
        {tagdc}
\bibfield{author}{\bibinfo{person}{Can Li}, \bibinfo{person}{Ling Xu},
  \bibinfo{person}{Meng Yan}, {and} \bibinfo{person}{Yan Lei}.}
  \bibinfo{year}{2020}\natexlab{}.
\newblock \showarticletitle{TagDC: A tag recommendation method for software
  information sites with a combination of deep learning and collaborative
  filtering}.
\newblock \bibinfo{journal}{\emph{Journal of Systems and Software}}
  \bibinfo{volume}{170} (\bibinfo{date}{08} \bibinfo{year}{2020}),
  \bibinfo{pages}{110783}.
\newblock
\urldef\tempurl%
\url{https://doi.org/10.1016/j.jss.2020.110783}
\showDOI{\tempurl}


\bibitem[\protect\citeauthoryear{Lin, Liu, Zeng, Jiang, and Cleland-Huang}{Lin
  et~al\mbox{.}}{2021}]%
        {tracebert}
\bibfield{author}{\bibinfo{person}{Jinfeng Lin}, \bibinfo{person}{Yalin Liu},
  \bibinfo{person}{Qingkai Zeng}, \bibinfo{person}{Meng Jiang}, {and}
  \bibinfo{person}{Jane Cleland-Huang}.} \bibinfo{year}{2021}\natexlab{}.
\newblock \showarticletitle{Traceability transformed: Generating more accurate
  links with pre-trained BERT models}. In \bibinfo{booktitle}{\emph{2021
  IEEE/ACM 43rd International Conference on Software Engineering (ICSE)}}.
  IEEE, \bibinfo{pages}{324--335}.
\newblock


\bibitem[\protect\citeauthoryear{Liu, Ott, Goyal, Du, Joshi, Chen, Levy, Lewis,
  Zettlemoyer, and Stoyanov}{Liu et~al\mbox{.}}{2019}]%
        {roberta}
\bibfield{author}{\bibinfo{person}{Yinhan Liu}, \bibinfo{person}{Myle Ott},
  \bibinfo{person}{Naman Goyal}, \bibinfo{person}{Jingfei Du},
  \bibinfo{person}{Mandar Joshi}, \bibinfo{person}{Danqi Chen},
  \bibinfo{person}{Omer Levy}, \bibinfo{person}{Mike Lewis},
  \bibinfo{person}{Luke Zettlemoyer}, {and} \bibinfo{person}{Veselin
  Stoyanov}.} \bibinfo{year}{2019}\natexlab{}.
\newblock \showarticletitle{Roberta: A robustly optimized bert pretraining
  approach}.
\newblock \bibinfo{journal}{\emph{arXiv preprint arXiv:1907.11692}}
  (\bibinfo{year}{2019}).
\newblock


\bibitem[\protect\citeauthoryear{Mashhadi and Hemmati}{Mashhadi and
  Hemmati}{2021}]%
        {mashhadi2021apply}
\bibfield{author}{\bibinfo{person}{Ehsan Mashhadi} {and} \bibinfo{person}{Hadi
  Hemmati}.} \bibinfo{year}{2021}\natexlab{}.
\newblock \showarticletitle{Applying CodeBERT for Automated Program Repair of
  Java Simple Bugs}. In \bibinfo{booktitle}{\emph{2021 IEEE/ACM 18th
  International Conference on Mining Software Repositories (MSR)}}.
  \bibinfo{pages}{505--509}.
\newblock
\urldef\tempurl%
\url{https://doi.org/10.1109/MSR52588.2021.00063}
\showDOI{\tempurl}


\bibitem[\protect\citeauthoryear{Ortu, Destefanis, Murgia, Tonelli, Marchesi,
  and Adams}{Ortu et~al\mbox{.}}{2015}]%
        {jira}
\bibfield{author}{\bibinfo{person}{Marco Ortu}, \bibinfo{person}{Giuseppe
  Destefanis}, \bibinfo{person}{Alessandro Murgia}, \bibinfo{person}{Roberto
  Tonelli}, \bibinfo{person}{Michele Marchesi}, {and} \bibinfo{person}{Bram
  Adams}.} \bibinfo{year}{2015}\natexlab{}.
\newblock \showarticletitle{The JIRA Repository Dataset: Understanding Social
  Aspects of Software Development}.
\newblock


\bibitem[\protect\citeauthoryear{Qiu, Sun, Xu, Shao, Dai, and Huang}{Qiu
  et~al\mbox{.}}{2020}]%
        {bertsurvey}
\bibfield{author}{\bibinfo{person}{Xipeng Qiu}, \bibinfo{person}{Tianxiang
  Sun}, \bibinfo{person}{Yige Xu}, \bibinfo{person}{Yunfan Shao},
  \bibinfo{person}{Ning Dai}, {and} \bibinfo{person}{Xuanjing Huang}.}
  \bibinfo{year}{2020}\natexlab{}.
\newblock \showarticletitle{Pre-trained models for natural language processing:
  A survey}.
\newblock \bibinfo{journal}{\emph{Science China Technological Sciences}}
  (\bibinfo{year}{2020}), \bibinfo{pages}{1--26}.
\newblock


\bibitem[\protect\citeauthoryear{Qu, Yang, Qiu, Croft, Zhang, and Iyyer}{Qu
  et~al\mbox{.}}{2019}]%
        {qu2019bert}
\bibfield{author}{\bibinfo{person}{Chen Qu}, \bibinfo{person}{Liu Yang},
  \bibinfo{person}{Minghui Qiu}, \bibinfo{person}{W~Bruce Croft},
  \bibinfo{person}{Yongfeng Zhang}, {and} \bibinfo{person}{Mohit Iyyer}.}
  \bibinfo{year}{2019}\natexlab{}.
\newblock \showarticletitle{BERT with history answer embedding for
  conversational question answering}. In \bibinfo{booktitle}{\emph{Proceedings
  of the 42nd international ACM SIGIR conference on research and development in
  information retrieval}}. \bibinfo{pages}{1133--1136}.
\newblock


\bibitem[\protect\citeauthoryear{Reimers and Gurevych}{Reimers and
  Gurevych}{2019}]%
        {reimers2019sentencebert}
\bibfield{author}{\bibinfo{person}{Nils Reimers} {and} \bibinfo{person}{Iryna
  Gurevych}.} \bibinfo{year}{2019}\natexlab{}.
\newblock \bibinfo{title}{Sentence-BERT: Sentence Embeddings using Siamese
  BERT-Networks}.
\newblock
\newblock
\showeprint[arxiv]{1908.10084}~[cs.CL]


\bibitem[\protect\citeauthoryear{Schmidhuber}{Schmidhuber}{2015}]%
        {schmidhuber2015deep}
\bibfield{author}{\bibinfo{person}{J{\"u}rgen Schmidhuber}.}
  \bibinfo{year}{2015}\natexlab{}.
\newblock \showarticletitle{Deep learning in neural networks: An overview}.
\newblock \bibinfo{journal}{\emph{Neural networks}}  \bibinfo{volume}{61}
  (\bibinfo{year}{2015}), \bibinfo{pages}{85--117}.
\newblock


\bibitem[\protect\citeauthoryear{Shi, Yang, He, Xu, and Lo}{Shi
  et~al\mbox{.}}{2022}]%
        {shi2022identifier}
\bibfield{author}{\bibinfo{person}{Jieke Shi}, \bibinfo{person}{Zhou Yang},
  \bibinfo{person}{Junda He}, \bibinfo{person}{Bowen Xu}, {and}
  \bibinfo{person}{David Lo}.} \bibinfo{year}{2022}\natexlab{}.
\newblock \showarticletitle{Can Identifier Splitting Improve Open-Vocabulary
  Language Model of Code?}. In \bibinfo{booktitle}{\emph{2022 IEEE
  International Conference on Software Analysis, Evolution and Reengineering
  (SANER)}}. \bibinfo{publisher}{IEEE}.
\newblock


\bibitem[\protect\citeauthoryear{Sun, Qiu, Xu, and Huang}{Sun
  et~al\mbox{.}}{2020}]%
        {howtofineture}
\bibfield{author}{\bibinfo{person}{Chi Sun}, \bibinfo{person}{Xipeng Qiu},
  \bibinfo{person}{Yige Xu}, {and} \bibinfo{person}{Xuanjing Huang}.}
  \bibinfo{year}{2020}\natexlab{}.
\newblock \bibinfo{title}{How to Fine-Tune BERT for Text Classification?}
\newblock
\newblock
\showeprint[arxiv]{1905.05583}~[cs.CL]


\bibitem[\protect\citeauthoryear{Svyatkovskiy, Deng, Fu, and
  Sundaresan}{Svyatkovskiy et~al\mbox{.}}{2020}]%
        {gpt-c}
\bibfield{author}{\bibinfo{person}{Alexey Svyatkovskiy},
  \bibinfo{person}{Shao~Kun Deng}, \bibinfo{person}{Shengyu Fu}, {and}
  \bibinfo{person}{Neel Sundaresan}.} \bibinfo{year}{2020}\natexlab{}.
\newblock \bibinfo{title}{IntelliCode Compose: Code Generation Using
  Transformer}.
\newblock
\newblock
\showeprint[arxiv]{2005.08025}~[cs.CL]


\bibitem[\protect\citeauthoryear{Tabassum, Maddela, Xu, and Ritter}{Tabassum
  et~al\mbox{.}}{2020}]%
        {bertoverflow}
\bibfield{author}{\bibinfo{person}{Jeniya Tabassum}, \bibinfo{person}{Mounica
  Maddela}, \bibinfo{person}{Wei Xu}, {and} \bibinfo{person}{Alan Ritter}.}
  \bibinfo{year}{2020}\natexlab{}.
\newblock \bibinfo{title}{Code and Named Entity Recognition in StackOverflow}.
\newblock
\newblock
\showeprint[arxiv]{2005.01634}~[cs.CL]


\bibitem[\protect\citeauthoryear{Vaswani, Shazeer, Parmar, Uszkoreit, Jones,
  Gomez, Kaiser, and Polosukhin}{Vaswani et~al\mbox{.}}{2017}]%
        {transformer}
\bibfield{author}{\bibinfo{person}{Ashish Vaswani}, \bibinfo{person}{Noam
  Shazeer}, \bibinfo{person}{Niki Parmar}, \bibinfo{person}{Jakob Uszkoreit},
  \bibinfo{person}{Llion Jones}, \bibinfo{person}{Aidan~N Gomez},
  \bibinfo{person}{{\L}ukasz Kaiser}, {and} \bibinfo{person}{Illia
  Polosukhin}.} \bibinfo{year}{2017}\natexlab{}.
\newblock \showarticletitle{Attention is all you need}. In
  \bibinfo{booktitle}{\emph{Advances in neural information processing
  systems}}. \bibinfo{pages}{5998--6008}.
\newblock


\bibitem[\protect\citeauthoryear{von~der Mosel, Trautsch, and Herbold}{von~der
  Mosel et~al\mbox{.}}{2021}]%
        {sebert}
\bibfield{author}{\bibinfo{person}{Julian von~der Mosel},
  \bibinfo{person}{Alexander Trautsch}, {and} \bibinfo{person}{Steffen
  Herbold}.} \bibinfo{year}{2021}\natexlab{}.
\newblock \showarticletitle{On the validity of pre-trained transformers for
  natural language processing in the software engineering domain}.
\newblock \bibinfo{journal}{\emph{CoRR}}  \bibinfo{volume}{abs/2109.04738}
  (\bibinfo{year}{2021}).
\newblock
\showeprint[arXiv]{2109.04738}


\bibitem[\protect\citeauthoryear{Wang, Lo, Vasilescu, and Serebrenik}{Wang
  et~al\mbox{.}}{2014}]%
        {entagrec}
\bibfield{author}{\bibinfo{person}{Shaowei Wang}, \bibinfo{person}{David Lo},
  \bibinfo{person}{Bogdan Vasilescu}, {and} \bibinfo{person}{Alexander
  Serebrenik}.} \bibinfo{year}{2014}\natexlab{}.
\newblock \showarticletitle{EnTagRec: An Enhanced Tag Recommendation System for
  Software Information Sites}. In \bibinfo{booktitle}{\emph{2014 IEEE
  International Conference on Software Maintenance and Evolution}}.
  \bibinfo{pages}{291--300}.
\newblock
\urldef\tempurl%
\url{https://doi.org/10.1109/ICSME.2014.51}
\showDOI{\tempurl}


\bibitem[\protect\citeauthoryear{Wang, Lo, Vasilescu, and Serebrenik}{Wang
  et~al\mbox{.}}{2018a}]%
        {entagresplusplus}
\bibfield{author}{\bibinfo{person}{Shaowei Wang}, \bibinfo{person}{David Lo},
  \bibinfo{person}{Bogdan Vasilescu}, {and} \bibinfo{person}{Alexander
  Serebrenik}.} \bibinfo{year}{2018}\natexlab{a}.
\newblock \showarticletitle{EnTagRec ++: An enhanced tag recommendation system
  for software information sites}.
\newblock \bibinfo{journal}{\emph{Empirical Software Engineering}}
  \bibinfo{volume}{23} (\bibinfo{date}{04} \bibinfo{year}{2018}).
\newblock


\bibitem[\protect\citeauthoryear{Wang, Lo, Vasilescu, and Serebrenik}{Wang
  et~al\mbox{.}}{2018b}]%
        {wang2018entagrec}
\bibfield{author}{\bibinfo{person}{Shaowei Wang}, \bibinfo{person}{David Lo},
  \bibinfo{person}{Bogdan Vasilescu}, {and} \bibinfo{person}{Alexander
  Serebrenik}.} \bibinfo{year}{2018}\natexlab{b}.
\newblock \showarticletitle{EnTagRec++: An enhanced tag recommendation system
  for software information sites}.
\newblock \bibinfo{journal}{\emph{Empirical Software Engineering}}
  \bibinfo{volume}{23}, \bibinfo{number}{2} (\bibinfo{year}{2018}),
  \bibinfo{pages}{800--832}.
\newblock


\bibitem[\protect\citeauthoryear{Wang, Xia, and Lo}{Wang
  et~al\mbox{.}}{2015a}]%
        {tagcombine}
\bibfield{author}{\bibinfo{person}{Xin-Yu Wang}, \bibinfo{person}{Xin Xia},
  {and} \bibinfo{person}{David Lo}.} \bibinfo{year}{2015}\natexlab{a}.
\newblock \showarticletitle{Tagcombine: Recommending tags to contents in
  software information sites}.
\newblock \bibinfo{journal}{\emph{Journal of Computer Science and Technology}}
  \bibinfo{volume}{30}, \bibinfo{number}{5} (\bibinfo{year}{2015}),
  \bibinfo{pages}{1017--1035}.
\newblock


\bibitem[\protect\citeauthoryear{Wang, Xia, and Lo}{Wang
  et~al\mbox{.}}{2015b}]%
        {wang2015tagcombine}
\bibfield{author}{\bibinfo{person}{Xin-Yu Wang}, \bibinfo{person}{Xin Xia},
  {and} \bibinfo{person}{David Lo}.} \bibinfo{year}{2015}\natexlab{b}.
\newblock \showarticletitle{Tagcombine: Recommending tags to contents in
  software information sites}.
\newblock \bibinfo{journal}{\emph{Journal of Computer Science and Technology}}
  \bibinfo{volume}{30}, \bibinfo{number}{5} (\bibinfo{year}{2015}),
  \bibinfo{pages}{1017--1035}.
\newblock


\bibitem[\protect\citeauthoryear{Williams, Nangia, and Bowman}{Williams
  et~al\mbox{.}}{2018}]%
        {williams-etal-2018-broad}
\bibfield{author}{\bibinfo{person}{Adina Williams}, \bibinfo{person}{Nikita
  Nangia}, {and} \bibinfo{person}{Samuel Bowman}.}
  \bibinfo{year}{2018}\natexlab{}.
\newblock \showarticletitle{A Broad-Coverage Challenge Corpus for Sentence
  Understanding through Inference}. In \bibinfo{booktitle}{\emph{Proceedings of
  the 2018 Conference of the North {A}merican Chapter of the Association for
  Computational Linguistics: Human Language Technologies, Volume 1 (Long
  Papers)}}. \bibinfo{publisher}{Association for Computational Linguistics},
  \bibinfo{address}{New Orleans, Louisiana}, \bibinfo{pages}{1112--1122}.
\newblock
\urldef\tempurl%
\url{https://doi.org/10.18653/v1/N18-1101}
\showDOI{\tempurl}


\bibitem[\protect\citeauthoryear{Xia, Lo, Wang, and Zhou}{Xia
  et~al\mbox{.}}{2013a}]%
        {xiaxin2013}
\bibfield{author}{\bibinfo{person}{Xin Xia}, \bibinfo{person}{David Lo},
  \bibinfo{person}{Xinyu Wang}, {and} \bibinfo{person}{Bo Zhou}.}
  \bibinfo{year}{2013}\natexlab{a}.
\newblock \showarticletitle{Tag Recommendation in Software Information Sites}.
  In \bibinfo{booktitle}{\emph{Proceedings of the 10th Working Conference on
  Mining Software Repositories}} (San Francisco, CA, USA)
  \emph{(\bibinfo{series}{MSR '13})}. \bibinfo{publisher}{IEEE Press},
  \bibinfo{pages}{287–296}.
\newblock
\showISBNx{9781467329361}


\bibitem[\protect\citeauthoryear{Xia, Lo, Wang, and Zhou}{Xia
  et~al\mbox{.}}{2013b}]%
        {xia2013tag}
\bibfield{author}{\bibinfo{person}{Xin Xia}, \bibinfo{person}{David Lo},
  \bibinfo{person}{Xinyu Wang}, {and} \bibinfo{person}{Bo Zhou}.}
  \bibinfo{year}{2013}\natexlab{b}.
\newblock \showarticletitle{Tag recommendation in software information sites}.
  In \bibinfo{booktitle}{\emph{2013 10th Working Conference on Mining Software
  Repositories (MSR)}}. IEEE, \bibinfo{pages}{287--296}.
\newblock


\bibitem[\protect\citeauthoryear{Xu, Hoang, Sharma, Yang, Xia, and Lo}{Xu
  et~al\mbox{.}}{2021}]%
        {post2vec}
\bibfield{author}{\bibinfo{person}{Bowen Xu}, \bibinfo{person}{Thong Hoang},
  \bibinfo{person}{Abhishek Sharma}, \bibinfo{person}{Chengran Yang},
  \bibinfo{person}{Xin Xia}, {and} \bibinfo{person}{David Lo}.}
  \bibinfo{year}{2021}\natexlab{}.
\newblock \showarticletitle{Post2Vec: Learning Distributed Representations of
  Stack Overflow Posts}.
\newblock \bibinfo{journal}{\emph{IEEE Transactions on Software Engineering}}
  (\bibinfo{year}{2021}), \bibinfo{pages}{1--1}.
\newblock
\urldef\tempurl%
\url{https://doi.org/10.1109/TSE.2021.3093761}
\showDOI{\tempurl}


\bibitem[\protect\citeauthoryear{Yang, Xu, Khan~Younus, Uddin, Han, Yang, and
  Lo}{Yang et~al\mbox{.}}{2022}]%
        {chengran2022saner}
\bibfield{author}{\bibinfo{person}{Chengran Yang}, \bibinfo{person}{Bowen Xu},
  \bibinfo{person}{Junaed Khan~Younus}, \bibinfo{person}{Gias Uddin},
  \bibinfo{person}{Donggyun Han}, \bibinfo{person}{Zhou Yang}, {and}
  \bibinfo{person}{David Lo}.} \bibinfo{year}{2022}\natexlab{}.
\newblock \showarticletitle{Aspect-Based API Review Classification: How Far Can
  Pre-Trained Transformer Model Go?}. In \bibinfo{booktitle}{\emph{29th IEEE
  International Conference onSoftware Analysis, Evolution and
  Reengineering(SANER)}}. IEEE.
\newblock


\bibitem[\protect\citeauthoryear{Zhang, Xu, Thung, Haryono, Lo, and
  Jiang}{Zhang et~al\mbox{.}}{2020}]%
        {zhang2020sentiment}
\bibfield{author}{\bibinfo{person}{Ting Zhang}, \bibinfo{person}{Bowen Xu},
  \bibinfo{person}{Ferdian Thung}, \bibinfo{person}{Stefanus~Agus Haryono},
  \bibinfo{person}{David Lo}, {and} \bibinfo{person}{Lingxiao Jiang}.}
  \bibinfo{year}{2020}\natexlab{}.
\newblock \showarticletitle{Sentiment analysis for software engineering: How
  far can pre-trained transformer models go?}. In
  \bibinfo{booktitle}{\emph{2020 IEEE International Conference on Software
  Maintenance and Evolution (ICSME)}}. IEEE, \bibinfo{pages}{70--80}.
\newblock


\bibitem[\protect\citeauthoryear{Zhou, Liu, Liu, Yang, and Grundy}{Zhou
  et~al\mbox{.}}{2019}]%
        {tagcnn}
\bibfield{author}{\bibinfo{person}{Pingyi Zhou}, \bibinfo{person}{Jin Liu},
  \bibinfo{person}{Xiao Liu}, \bibinfo{person}{Zijiang Yang}, {and}
  \bibinfo{person}{John Grundy}.} \bibinfo{year}{2019}\natexlab{}.
\newblock \showarticletitle{Is deep learning better than traditional approaches
  in tag recommendation for software information sites?}
\newblock \bibinfo{journal}{\emph{Information and Software Technology}}
  \bibinfo{volume}{109} (\bibinfo{date}{May} \bibinfo{year}{2019}),
  \bibinfo{pages}{1--13}.
\newblock
\showISSN{0950-5849}
\urldef\tempurl%
\url{https://doi.org/10.1016/j.infsof.2019.01.002}
\showDOI{\tempurl}


\bibitem[\protect\citeauthoryear{Zhou, Liu, Yang, and Zhou}{Zhou
  et~al\mbox{.}}{2017}]%
        {tagmulrec}
\bibfield{author}{\bibinfo{person}{Pingyi Zhou}, \bibinfo{person}{Jin Liu},
  \bibinfo{person}{Zijiang Yang}, {and} \bibinfo{person}{Guangyou Zhou}.}
  \bibinfo{year}{2017}\natexlab{}.
\newblock \showarticletitle{Scalable tag recommendation for software
  information sites}. In \bibinfo{booktitle}{\emph{2017 IEEE 24th International
  Conference on Software Analysis, Evolution and Reengineering (SANER)}}. IEEE,
  \bibinfo{pages}{272--282}.
\newblock


\bibitem[\protect\citeauthoryear{Zhou, Han, and Lo}{Zhou et~al\mbox{.}}{2021}]%
        {zhou2021assessing}
\bibfield{author}{\bibinfo{person}{Xin Zhou}, \bibinfo{person}{DongGyun Han},
  {and} \bibinfo{person}{David Lo}.} \bibinfo{year}{2021}\natexlab{}.
\newblock \showarticletitle{Assessing Generalizability of CodeBERT}. In
  \bibinfo{booktitle}{\emph{2021 IEEE International Conference on Software
  Maintenance and Evolution (ICSME)}}. IEEE, \bibinfo{pages}{425--436}.
\newblock


\bibitem[\protect\citeauthoryear{Zhu, Kiros, Zemel, Salakhutdinov, Urtasun,
  Torralba, and Fidler}{Zhu et~al\mbox{.}}{2015}]%
        {Zhu_2015_ICCV}
\bibfield{author}{\bibinfo{person}{Yukun Zhu}, \bibinfo{person}{Ryan Kiros},
  \bibinfo{person}{Rich Zemel}, \bibinfo{person}{Ruslan Salakhutdinov},
  \bibinfo{person}{Raquel Urtasun}, \bibinfo{person}{Antonio Torralba}, {and}
  \bibinfo{person}{Sanja Fidler}.} \bibinfo{year}{2015}\natexlab{}.
\newblock \showarticletitle{Aligning Books and Movies: Towards Story-Like
  Visual Explanations by Watching Movies and Reading Books}. In
  \bibinfo{booktitle}{\emph{The IEEE International Conference on Computer
  Vision (ICCV)}}.
\newblock


\end{thebibliography}
\end{document}